\documentclass[preprint]{vgtc}               %

\ifpdf%
  \pdfoutput=1\relax                   %
  \usepackage{graphicx}                %
  \DeclareGraphicsExtensions{.pdf,.png,.jpg,.jpeg} %
\else%
  \usepackage{graphicx}                %
  \DeclareGraphicsExtensions{.eps}     %
\fi%

\graphicspath{{figures/}{pictures/}{images/}{./}} %

\usepackage{microtype}                 %
\PassOptionsToPackage{warn}{textcomp}  %
\usepackage{textcomp}                  %
\usepackage{mathptmx}                  %
\usepackage{times}                     %
\usepackage{cite}                      %
\usepackage{tabu}                      %
\usepackage{booktabs}                  %

\vgtcinsertpkg

\usepackage{subcaption}
\usepackage{enumitem}
\usepackage{tabularx}

\onlineid{1090}

\vgtccategory{System}

\title{MoPeDT: A Modular Head-Mounted Display Toolkit to Conduct Peripheral Vision Research}

\author{Matthias Albrecht\thanks{e-mail: matthias.albrecht@uni-konstanz.de}\\ %
        \scriptsize University of Konstanz, Germany %
\and Lorenz Assländer\thanks{e-mail: lorenz.asslaender@uni-konstanz.de}\\ %
     \scriptsize University of Konstanz, Germany %
\and Harald Reiterer\thanks{e-mail: harald.reiterer@uni-konstanz.de}\\ %
     \scriptsize University of Konstanz, Germany %
\and Stephan Streuber\thanks{e-mail: stephan.streuber@hs-coburg.de}\\ %
     \scriptsize Coburg University of Applied Sciences and Arts, Germany} %

\teaser{
    \centering
    \begin{subfigure}[t]{0.245\linewidth}
        \centering
        \includegraphics[width=\linewidth]{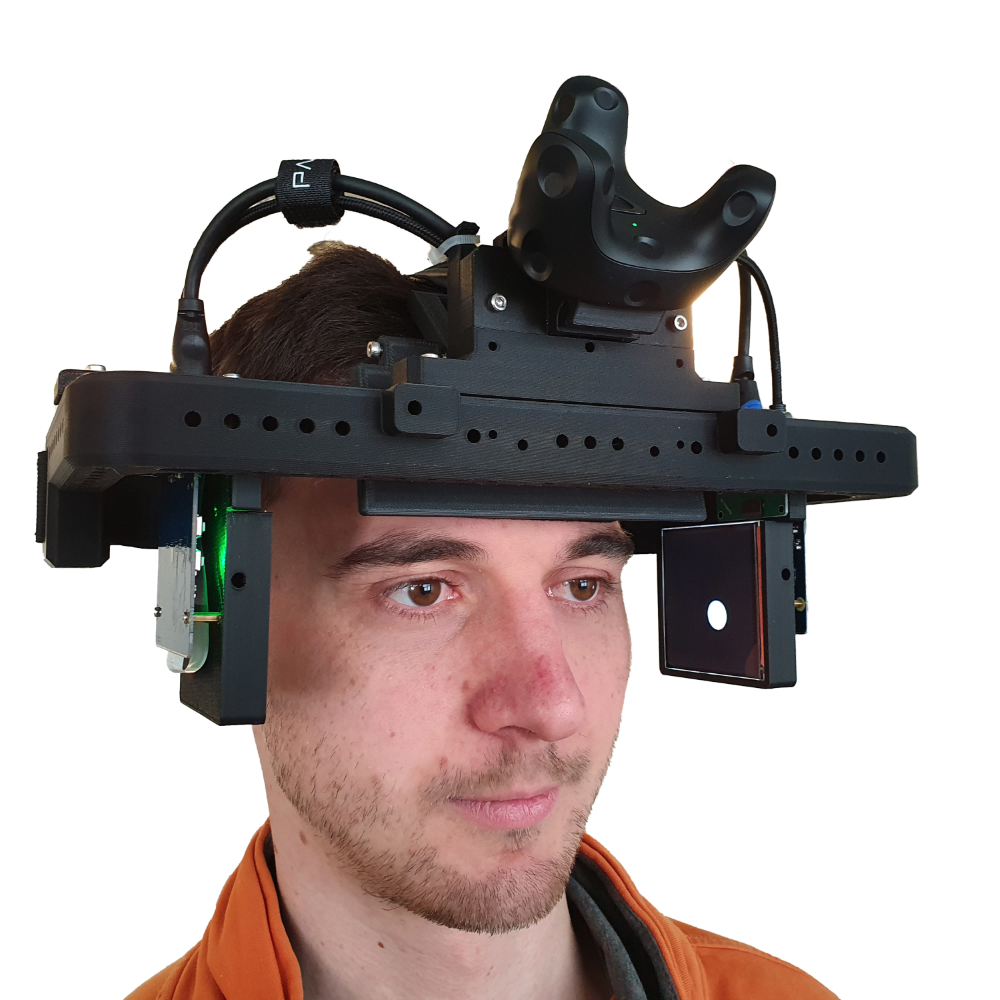}
        \caption{Parallel LCDs and HTC Vive tracker}
        \label{fig:teaser-configuration-parallel}
    \end{subfigure}
    \hfill
    \begin{subfigure}[t]{0.245\linewidth}
        \centering
        \includegraphics[width=\linewidth]{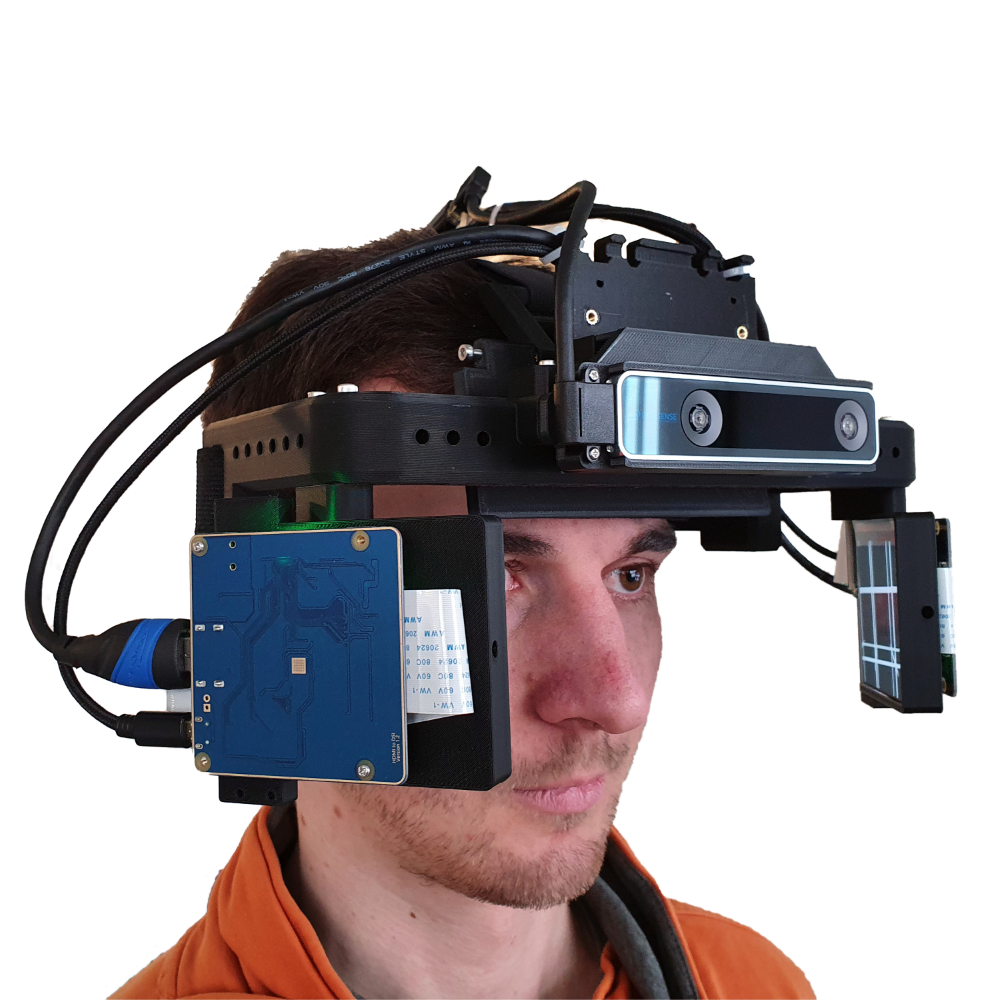}
        \caption{Angled LCDs and Intel RealSense T265 tracking camera}
        \label{fig:teaser-configuration-angled}
    \end{subfigure}
    \hfill
    \begin{subfigure}[t]{0.245\linewidth}
        \centering
        \includegraphics[width=\linewidth]{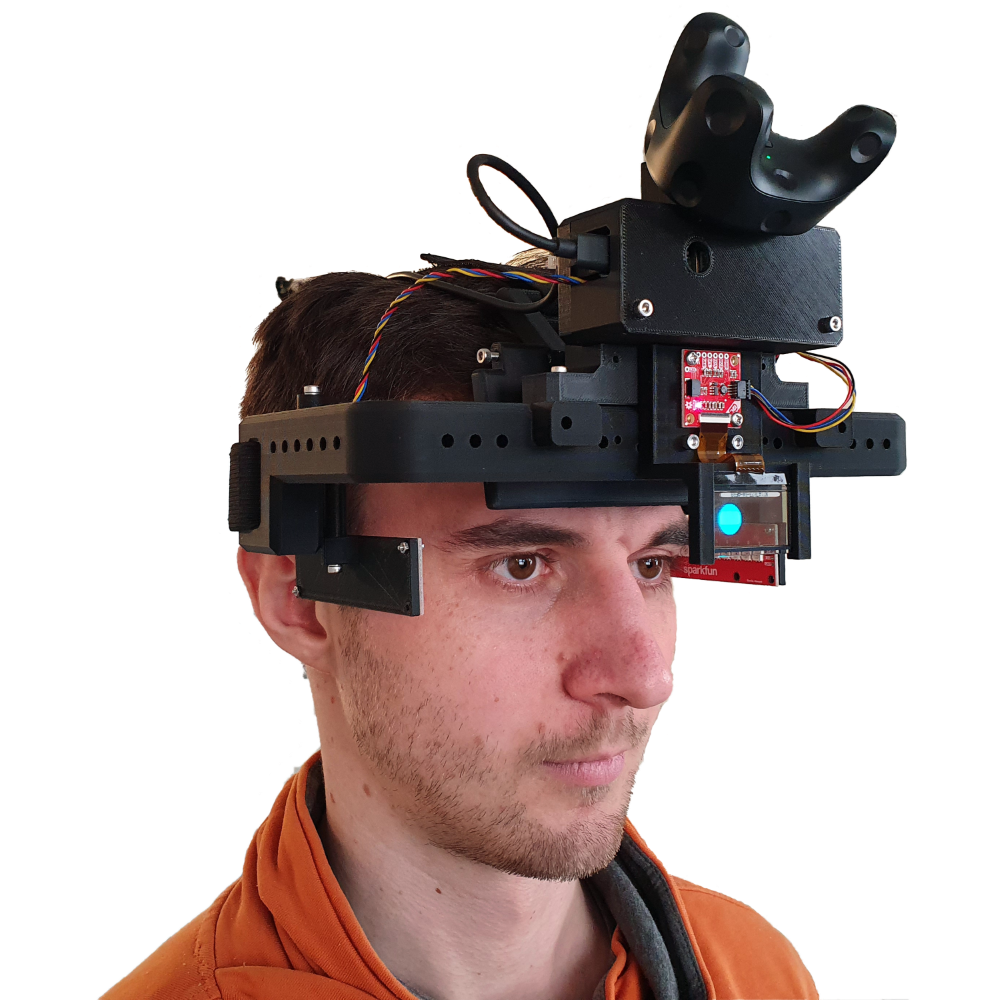}
        \caption{LED sticks, transparent OLED display and HTC Vive tracker}
        \label{fig:teaser-configuration-led-stick}
    \end{subfigure}
    \hfill
    \begin{subfigure}[t]{0.245\linewidth}
        \centering
        \includegraphics[width=\linewidth]{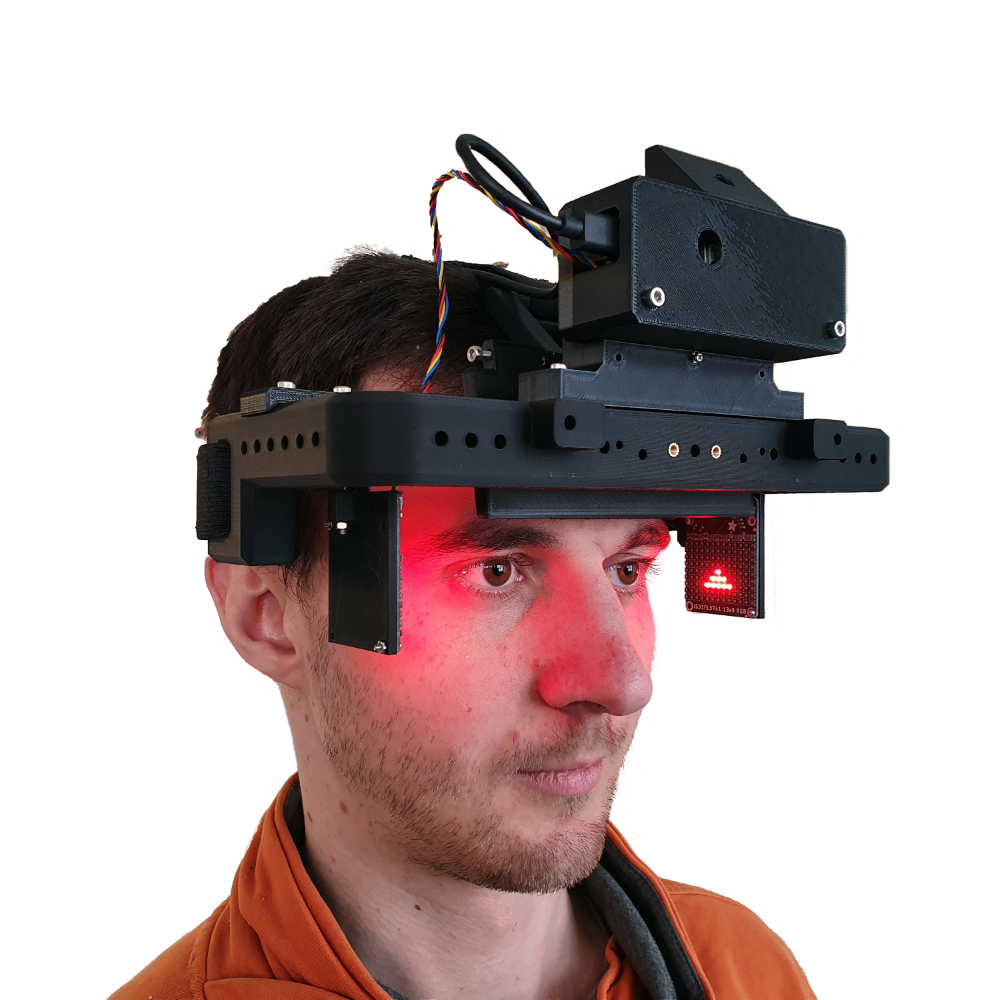}
        \caption{LED matrices without extra tracking module}
        \label{fig:teaser-configuration-led-matrix}
    \end{subfigure}
    \caption{A user wearing four possible configurations of MoPeDT with different display and tracking modules.}
    \label{fig:teaser-configurations}
}

\abstract{Peripheral vision plays a significant role in human perception and orientation. However, its relevance for human-computer interaction, especially head-mounted displays, has not been fully explored yet. In the past, a few specialized appliances were developed to display visual cues in the periphery, each designed for a single specific use case only. A multi-purpose headset to exclusively augment peripheral vision did not exist yet. We introduce MoPeDT: Modular Peripheral Display Toolkit, a freely available, flexible, reconfigurable, and extendable headset to conduct peripheral vision research. MoPeDT can be built with a 3D printer and off-the-shelf components. It features multiple spatially configurable near-eye display modules and full 3D tracking inside and outside the lab. With our system, researchers and designers may easily develop and prototype novel peripheral vision interaction and visualization techniques. We demonstrate the versatility of our headset with several possible applications for spatial awareness, balance, interaction, feedback, and notifications. We conducted a small study to evaluate the usability of the system. We found that participants were largely not irritated by the peripheral cues, but the headset's comfort could be further improved. We also evaluated our system based on established heuristics for human-computer interaction toolkits to show how MoPeDT adapts to changing requirements, lowers the entry barrier for peripheral vision research, and facilitates expressive power in the combination of modular building blocks.%
} %

\CCScatlist{
  \CCScatTwelve{Human-centered computing}{Human computer interaction (HCI)}{Interaction paradigms}{Mixed / augmented reality};
  \CCScatTwelve{Human-centered computing}{Human computer interaction (HCI)}{Interactive systems and tools}{User interface toolkits};
}

\newcommand{\citeauthor}[1]{\textcolor{red}{MISSING AUTHOR}}
\newcolumntype{L}{>{\raggedright\arraybackslash}X}

\begin{document}
\firstsection{Introduction}

\maketitle

Peripheral vision is a key aspect of human visual perception. It allows us to be aware of what is happening around us in the environment, warns us of imminent dangers, and aids orientation. The human field of view spans around 200 to 220 degrees horizontally and 130 to 150 degrees vertically, without accounting for additional eye movement \cite{strasburger_seven_2020}. It can be divided into central and peripheral regions, which have different properties with regard to what can be perceived. The periphery is especially weak in distinguishing detail but excels at discerning motion and contrasts \cite{strasburger_peripheral_2011}. It plays an important role in structure recognition, maintaining body posture, and spatial understanding \cite{palmer_vision_1999}.

Researching peripheral vision in augmented reality (AR), virtual reality (VR), and human-computer interaction (HCI) is hindered by a lack of easy-to-use hardware solutions for applications and experimentation. The existing studies primarily explore the application of peripheral vision for enhancing spatial awareness \cite{xiao_augmenting_2016,gruenefeld_radiallight_2018, gruenefeld_guiding_2018, tseng_lead_2015}, improving AR/VR experiences by reducing motion sickness or increasing presence \cite{xiao_augmenting_2016,hashemian_investigating_2018,nakano_head-mounted_2021}, and notifications \cite{costanza_eye-q_2006, luyten_hidden_2016}. Those studies show promising results on the use of peripheral displays for AR/VR and HCI. Due to the absence of existing commercially available devices that support peripheral-only visual cues, specialized headsets have been developed. Those devices work well for their advertised use cases but are difficult to repurpose in order to develop and test new peripheral vision concepts. Many approaches solely use LED-based visual stimuli that are inherently limited in terms of what can be displayed. They often also do not feature tracking, as it is common with modern AR/VR head-mounted displays (HMD) and necessary to render content that is aware of the user’s head position and orientation. Some VR headsets offer a wide field of view of 100 degrees and above but they isolate the user from their real environment and as such cannot be used to investigate peripheral AR. Due to these limitations, such devices cannot be easily adapted or extended for other uses beyond what they were designed for. Also, building and replicating such a headset from the publication alone is challenging without a bill of materials or instructions. Unlike many available traditional AR and VR devices, a multi-purpose HMD specifically tailored for peripheral vision did not exist yet. As a result of these hurdles, we believe that many promising peripheral visualization or interaction techniques may not have been explored yet. 

Here we propose a new way for researchers to prototype and test novel peripheral vision applications using near-eye displays by introducing a modular, flexible and extendable headset: \emph{MoPeDT - Modular Peripheral Display Toolkit}. MoPeDT is designed to exclusively augment only the peripheral vision and leaves central vision free of obstructions. Conventional AR or mixed reality (MR) glasses allow the wearer to perceive and participate in the real world, but since their field of view is relatively small they can only supplement a limited area with additional visual information in mostly the foveal view. Exclusively augmenting the periphery with near-eye displays leaves the central visual region free. This is a promising concept for giving unobtrusive visual cues or notifications to the user and enhancing their senses without requiring an attention shift away from their current task at hand or obstructing central vision. This is especially important in situations where being distracted may lead to serious harm like in traffic.
By augmenting peripheral vision, we tap into an already existing channel of human perception.

MoPeDT is a standalone device and is currently not intended to be combined with other AR/VR HMDs. It features different LCD and LED display modules that can be freely attached to a head-mounted frame. To fulfill different requirements on how the headset will be used, it supports three tracking options that can work inside and outside a laboratory environment without necessarily depending on a fixed reference frame. Through easy reconfiguration, researchers can quickly implement and try out different designs without the need to create entirely new hardware. All resources necessary to 3D print and build the headset, including CAD files, component lists, and software, are freely available and open-source \cite{Albrecht_MoPeDT_2023}. To the best of our knowledge, MoPeDT is the first dedicated HMD platform that allows different arrangements of display elements to exclusively augment peripheral vision in combination with flexible tracking options. 

In this paper, we first review related works that use peripheral visual cues in the context of HMDs. Then, we describe the derived design requirements and implementation of our proposed system, detailing how it is built and what features it offers. Afterward, we motivate promising possible applications of peripheral-only cues such as improving spatial awareness, balance control, interaction, feedback, and notifications. To show the versatility and value of MoPeDT, we implemented those applications to showcase the different functionalities of our headset. Additionally, we present the subjective results of users wearing MoPeDT and a heuristic evaluation. Finally, we discuss our system, compare our approach to previous solutions, talk about limitations, and look toward future advancements.

\subsection{Contributions}
We present three main contributions:

\begin{enumerate}
    \item A modular and freely available headset for research that specifically targets peripheral visual cues. It comes with a set of software and hardware components that may enable new peripheral vision applications.
    \item Multiple demonstration applications that showcase the versatility of our toolkit through promising use cases of peripheral vision for AR, HCI, and other disciplines such as sports science.
    \item An evaluation of our system using the heuristics from Olsen \cite{olsen_evaluating_2007}, examining how MoPeDT can be adapted to new requirements and repurposed. We also present the results of a user study and compare our solution to related works.
\end{enumerate}

\section{Related Work}\label{sec:related-work}
In the second half of the 20th century, the aviation industry picked up peripheral vision to improve pilot performance for landing and orientation \cite{malcolm_pilot_1984,nasa_peripheral_1984}.
In AR/VR and HCI, peripheral visual perception has been the topic of many publications.
Shelton and Nesbitt performed a systematic review of multimodal ambient and peripheral information systems from 1996 to 2016 in a variety of usage contexts \cite{shelton_systematic_2020}. A vast majority (87\%) of the analyzed papers used visual information and around half of them are screen-based (49\%). Most of them either use physical lights, monitors on a desk, or projections. A big disadvantage of such devices is that they are stationary and do not follow the user. The setup is bound to the room or working space where it is installed.

\subsection{Extending HMDs with peripheral cues}
Some authors explored how to combine conventional AR and VR HMDs with additional peripheral cues. The benefit of augmenting existing AR/VR devices is that these already come with sophisticated tracking solutions and are generally easy to work with. 
Nakano et al. increased the downward field of view in VR with two additional screens placed below the eyes \cite{nakano_head-mounted_2021}. They were able to improve presence and sense of self-location with this configuration.
Xiao and Benko integrated LED arrays into an AR and VR headset to increase the perceivable field of view \cite{xiao_augmenting_2016}. They showed that these sparse visual cues were able to improve situational awareness and could help reduce motion sickness during usage. An extended version was developed by Hashemian et al. to study the effect of peripheral cues on self-movement and vection \cite{hashemian_investigating_2018}. They were able to show increased believability of motion in VR without increasing motion sickness.
Gruenefeld et al. proposed a similar technical approach to convey directional information and guide a wearer towards out-of-view objects using radially placed LEDs around the lenses of a VR HMD \cite{gruenefeld_radiallight_2018}. Their results suggest that visual cues in the periphery are suitable to guide users toward out-of-view objects. The device was based on the PeriMR prototyping toolkit to extend HMDs with peripheral light displays \cite{gruenefeld_perimr_2017}. Similar to MoPeDT, the authors follow an extendable and low-cost approach. However, PeriMR was only designed to feature LED-based peripheral cues and does not feature different display or tracking modules. Also, due to being based on Google Cardboard, a significant portion of the user's field of view is blocked which is something we want to avoid with MoPeDT.
Endo et al. presented an HMD concept that allows users ad-hoc peripheral interactions with their surroundings \cite{endo_modularhmd_2021}. Their system combines an HMD with removable modules in the periphery. While similar to MoPeDT in terms of modularity and placement of displays in the periphery, the authors follow a fundamentally different approach. The central HMD component is always present and the user can only see the real world when removing one of the side modules. With MoPeDT we want the wearer to perceive most of the real world and only augment selected parts in the periphery.

\subsection{Exclusively augmenting peripheral vision}
Due to advancements in 3D printing and ready-to-use electronics over the last years, head-worn specialized peripheral displays have been created.
Unlike those approaches that integrated peripheral cues for conventional HMDs, the following works exclusively augment the peripheral region with additional visual information. Hence, the user still has an unobstructed view of their environment in the central vision.

An early example of this was a wearable peripheral display that uses LEDs in the far periphery to show subtle moving visual cues for notifications \cite{costanza_eye-q_2006}. The authors found that these cues are effective in notifying the wearer. However, depending on the workload, some cues are suppressed and are less noticeable.
AmbiGlasses used multiple single LEDs attached around the frame of eyeglasses to display peripheral directional cues \cite{poppinga_ambiglasses_2012}. Users were able to estimate the rough location of a light stimulus with great accuracy. Similar eyeglasses using LEDs on the top of the frame were proposed by Olwal and Kress for pedestrian navigation \cite{olwal_1d_2018}. 
A related concept for peripheral visual guidance was introduced by Tseng et al. \cite{tseng_lead_2015}. They used an LED strip inside a scooter helmet, positioned in the upper peripheral vision field, to guide wearers with simple moving light patterns. A user study supports their claim that this is an effective way to direct scooter drivers.
Nakao et al. presented smart glasses with a peripherally attached 8x8 LED matrix and found that users were able to distinguish animation patterns with high accuracy \cite{nakao_smart_2016}. 
To help skiers gain better spatial awareness and decrease the risk of accidents, Niforatos et al. introduced a head-mounted collision detection system using LIDAR sensors \cite{niforatos_augmenting_2017}. Three peripheral LEDs inside a helmet warn the wearer about such an imminent event. Their study indicates that this system increases the peripheral perception of skiers.
Kiss et al. proposed Clairbuoyance, a pair of glasses using peripheral light stimuli to guide swimmers \cite{kiss_clairbuoyance_2019}. With the system, they could reach targets faster, improving orientation in the water.  
Cobus et al. suggested using similar peripheral light stimulation to give spatial alarms to caregivers in intensive care units \cite{cobus_multimodal_2017}.
Gruenefeld et al. developed two LED-based peripheral glasses to shift the attention of smartphone users and prevent accidents due to being distracted \cite{gruenefeld_guiding_2018}. In the first iteration, they used a simple horizontal strip of LEDs, then improved their design by incorporating an LED matrix to show a greater variety of visual cues. They found that moving light stimuli yielded the lowest response time and were subjectively preferred. By using radial peripheral light signals for visual guidance, van Veen et al. could increase situational awareness for drivers  \cite{van_veen_situation_2017}.

Luyten et al. explored the visual language that is suitable for displaying peripheral information by placing two small LCD screens very close to the left and right side of the eyes \cite{luyten_hidden_2016}. Based on a user study, they formulated a set of guidelines and recommendations on how to create peripheral visual cues that are easy to recognize. They suggest using simple prominent shapes from a limited predefined set, avoiding composite shapes, using few and strongly contrasting colors, and, most importantly, facilitating motion. We used these guidelines when creating the peripheral visualization for MoPeDT. A benefit of displaying content in a region where the eyes cannot directly focus is that no additional lenses are required to present a sharp image.

\section{System}
In this section, we will present our system: \emph{MoPeDT - Modular Peripheral Display Toolkit}. We introduce the requirements that guided the design and development process, describe the hardware, software and materials, illustrate the modular assembly process, and explain the device's functionality. 

\subsection{Design Requirements}
We compiled a set of requirements by looking at previous work that utilized peripheral cues. 
The devices shown so far that exclusively augment peripheral vision were designed for specific purposes (e.g., avoiding traffic accidents) or are generally limited to a few selected use cases \cite{costanza_eye-q_2006,poppinga_ambiglasses_2012,tseng_lead_2015,nakao_smart_2016,niforatos_augmenting_2017,kiss_clairbuoyance_2019,cobus_multimodal_2017,gruenefeld_guiding_2018,olwal_1d_2018,van_veen_situation_2017,luyten_hidden_2016}. With MoPeDT, we want to provide a more adaptable advancement of these approaches by proposing a unified and extendable system that allows a wide variety of applications. The headset should be modular to allow easy and flexible reconfiguration. During development and prototyping, many variations are tested and requirements may change. This is especially important for display elements to try out different spatial arrangements and viewing angles. Previous approaches offered only fixed positions of the display elements mounted to an eyeglass frame and did not allow fine-grained manipulation of the visual angle or eccentricity of peripheral cues which is important for peripheral vision research. Thus, our implementation needs a mechanism to control the position, distance, and rotation of display modules.
Some authors used LEGO bricks for rapid hardware prototyping and reconfiguration \cite{mueller_fabrickation_2014,orlosky_modular_2015}. In an early design phase, we also experimented with LEGO bricks to get a feel for the headset shape but we found that we need custom 3D printed modules, mainly due to rigidity, especially for the display and electronics components.
Unlike traditional AR glasses or VR HMDs that augment the central visual field or isolate the user entirely from the real world, we want to focus on the periphery only without the intent to display content in the foveal central visual field like conventional HMDs.

Other than those approaches that use a commercial AR or VR device \cite{nakano_head-mounted_2021,xiao_augmenting_2016,hashemian_investigating_2018,gruenefeld_radiallight_2018,gruenefeld_perimr_2017}, only a few dedicated peripheral HMDs are able to use rotation data from an IMU to base their visualization on the head orientation \cite{olwal_1d_2018,kiss_clairbuoyance_2019}. None of them can track their location in space. For MoPeDT, we want to overcome this limitation by integrating tracking solutions to enable more advanced peripheral applications that incorporate the position and attitude of the wearer. As MoPeDT is not primarily designed to be combined with other HMDs that already come with a tracking solution, it must provide tracking on its own.
In addition, tracking must also be available outside a controlled laboratory environment where no fixed tracking reference frame is available, e.g. when using the headset in the city.

Using just LEDs in the periphery limits the visual expressiveness due to the low number of light-emitting elements compared to higher resolution screens \cite{costanza_eye-q_2006,poppinga_ambiglasses_2012,tseng_lead_2015,nakao_smart_2016,niforatos_augmenting_2017,kiss_clairbuoyance_2019,cobus_multimodal_2017,gruenefeld_guiding_2018,olwal_1d_2018,van_veen_situation_2017}. Similar to Luyten et al. \cite{luyten_hidden_2016}, our system should support small near-eye LCD displays to allow greater visual expressiveness to display a wide variety of shapes, colors, and animations. In addition, LED-based display elements should also be available to use for simpler visualizations. We decided that more advanced display technologies like micro-projectors or waveguides are not suitable for the purpose of a prototyping headset due to their high cost, low availability, and technical difficulties.

The headset should be low-cost and easy to manufacture without requiring specialized machinery. Hardware parts should be off-the-shelf, ready-to-buy, and support a plug-and-play approach.
Creating software for our headset should leverage already available tools and frameworks that are not specific to our hardware to provide researchers with a straightforward way to implement specific custom peripheral vision applications and prototypes. 

While ergonomics are a very important aspect of a head-worn device, we wanted to follow a functional-first approach as the primary focus is on the perceptual exploration of peripheral cues and we do not expect that MoPeDT will be worn for an extended amount of time.
Therefore, ergonomics was not a key design requirement.

In summary, the identified design requirements are:
\begin{enumerate}
    \item Modular and flexible reconfiguration, multi-purpose, focus on augmenting the peripheral vision with visual cues.
    \item Full 3D tracking inside the lab and also outside in an uncontrolled environment.
    \item Small form-factor displays that are suitable for mounting to a head-worn device, capable of showing both low and high-fidelity content.
    \item Low-cost, easy-to-manufacture, off-the-shelf components. Plug-and-play.
    \item Use existing and well-known software frameworks and tools.
\end{enumerate}

\subsection{Implementation}\label{sec:implementation}
The headset consists of a frame as well as several independent display, tracking, and electronic processing modules. As modularity was a key design requirement, modules can be easily attached and replaced to create different configurations of the headset. All components are off-the-shelf electronics, screws, and threads or can be 3D printed. Users of MoPeDT may extend its functionality and create new components using standard CAD modeling software and 3D printers, which are affordable and already pretty commonly found in labs today.

Figure \ref{fig:teaser-configurations} and \ref{fig:configuration-ar} show different possible headset configurations using different modules.
Figure \ref{fig:frame} shows the basic 3D printed components that make up the base frame to which the modules are attached. 
See Figure \ref{fig:display-modules} for an overview of display and tracking modules that are available for MoPeDT.
Depending on the attached modules, the headset weighs between 350 to 580 g. 

\subsubsection{Head-worn frame}\label{sec:frame}
\begin{figure}
    \centering
    \includegraphics[width=\linewidth]{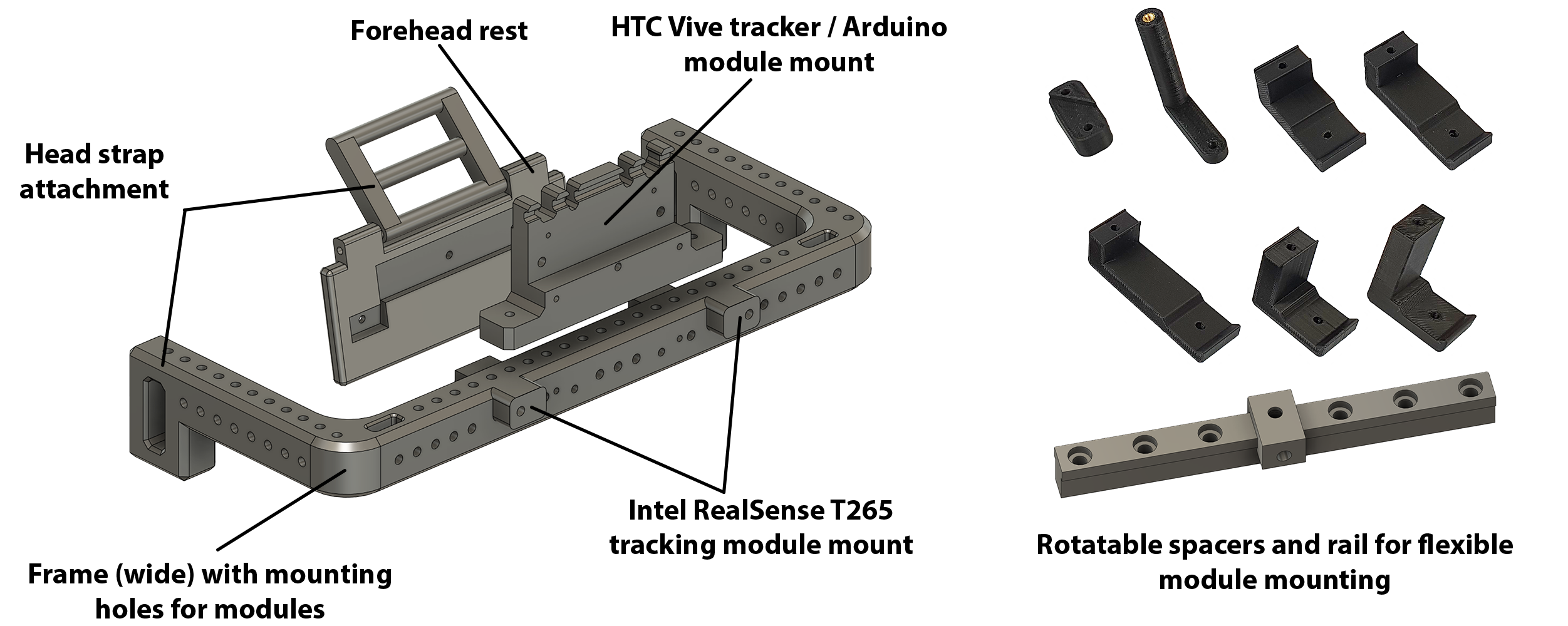}
    \caption{Base frame, rotatable spacers, and linear rail to attach modules.}
    \label{fig:frame}
\end{figure}

The central element of the headset is a sturdy U-shaped frame (Figure \ref{fig:frame}). It features numerous mounting holes for M3 heat-set inserts on each side, spaced 10 mm apart. This allows flexible assembly, especially during a design phase where the final position and orientation of the modules are not determined yet. Individual components are attached to the base frame using screws. The dimensions of the frame were chosen in a way so that usual human head sizes fit inside of it while leaving enough space for the placement of displays at the sides. The average human head breadth is around 146 mm for women and 152 mm for men (standard deviation $\approx\pm5$ mm for both) \cite{young_head_1993}. We offer two frame sizes, 250 and 300 mm wide (inner diameter 210 and 260 mm respectively), to accommodate different head sizes. The larger one features additional mounting holes and more space when using wider display modules. 
The components were printed on an Original Prusa i3 MK3S and Raise3D Pro2 Plus using polylactide filament (PLA).
An HTC Vive head strap is used to firmly secure the headset on the head, pressing against the forehead using a slightly angled headrest. A Velcro cushion increases comfort for the wearer.

\subsubsection{Display modules}

\begin{figure}
    \centering
    \begin{subfigure}[t]{0.19\linewidth}
        \centering
        \includegraphics[width=\linewidth]{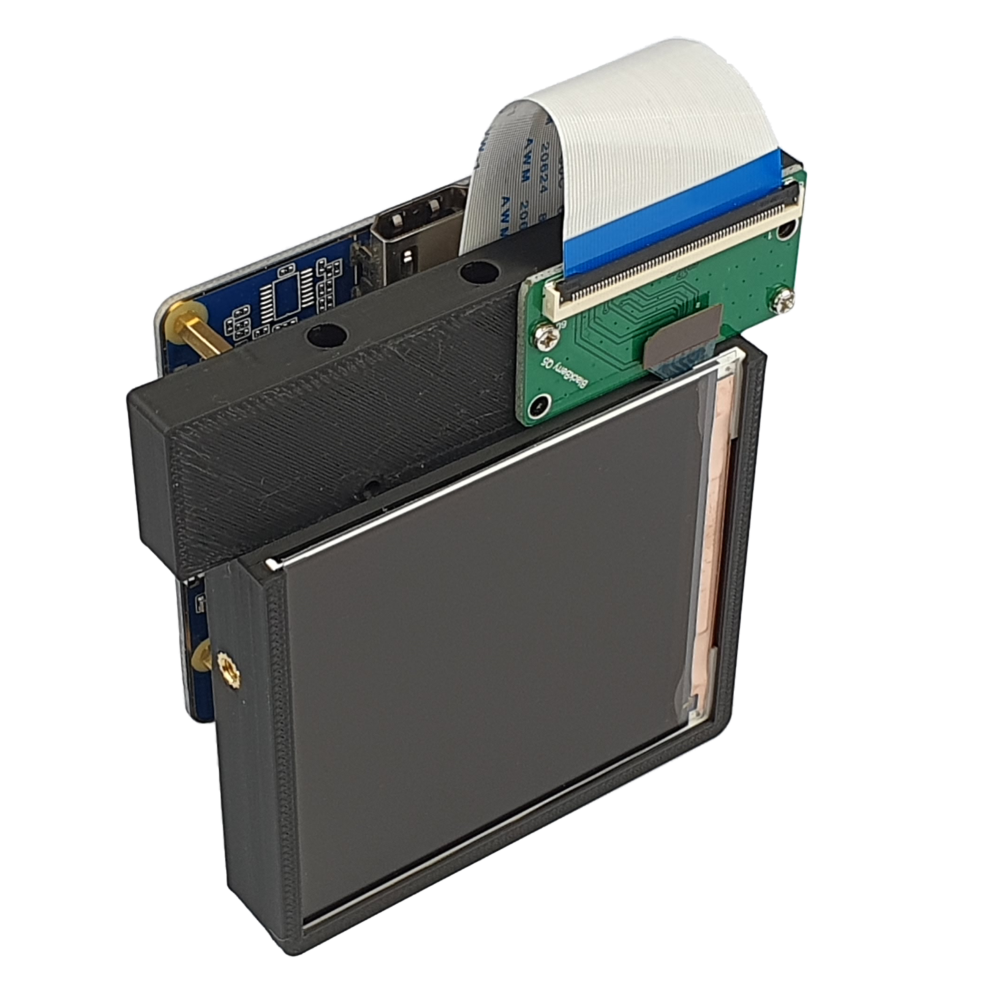}
        \caption{LCD display}
        \label{fig:module-lcd}
    \end{subfigure}
    \hfill
        \begin{subfigure}[t]{0.19\linewidth}
        \centering
        \includegraphics[width=\linewidth]{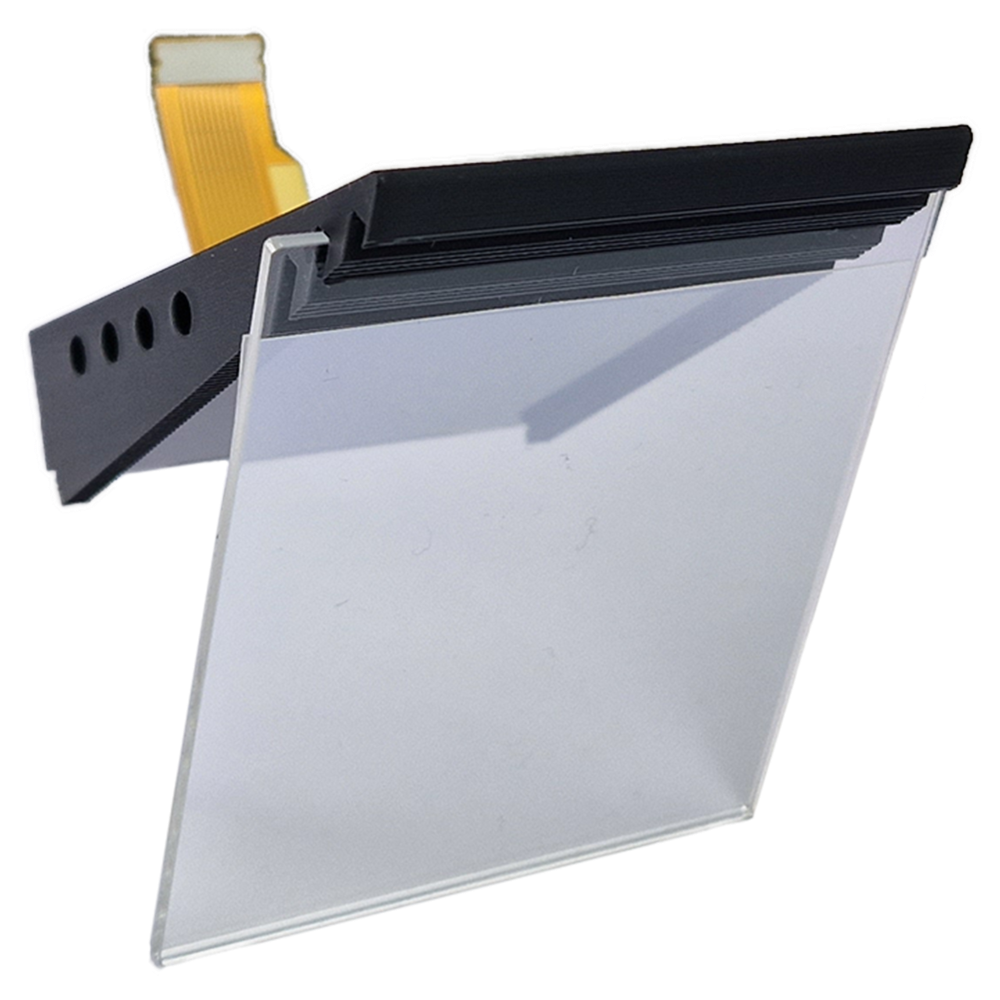}
        \caption{See-through AR module}
        \label{fig:module-ar}
    \end{subfigure}
    \hfill
    \begin{subfigure}[t]{0.19\linewidth}
        \centering
        \includegraphics[width=\linewidth]{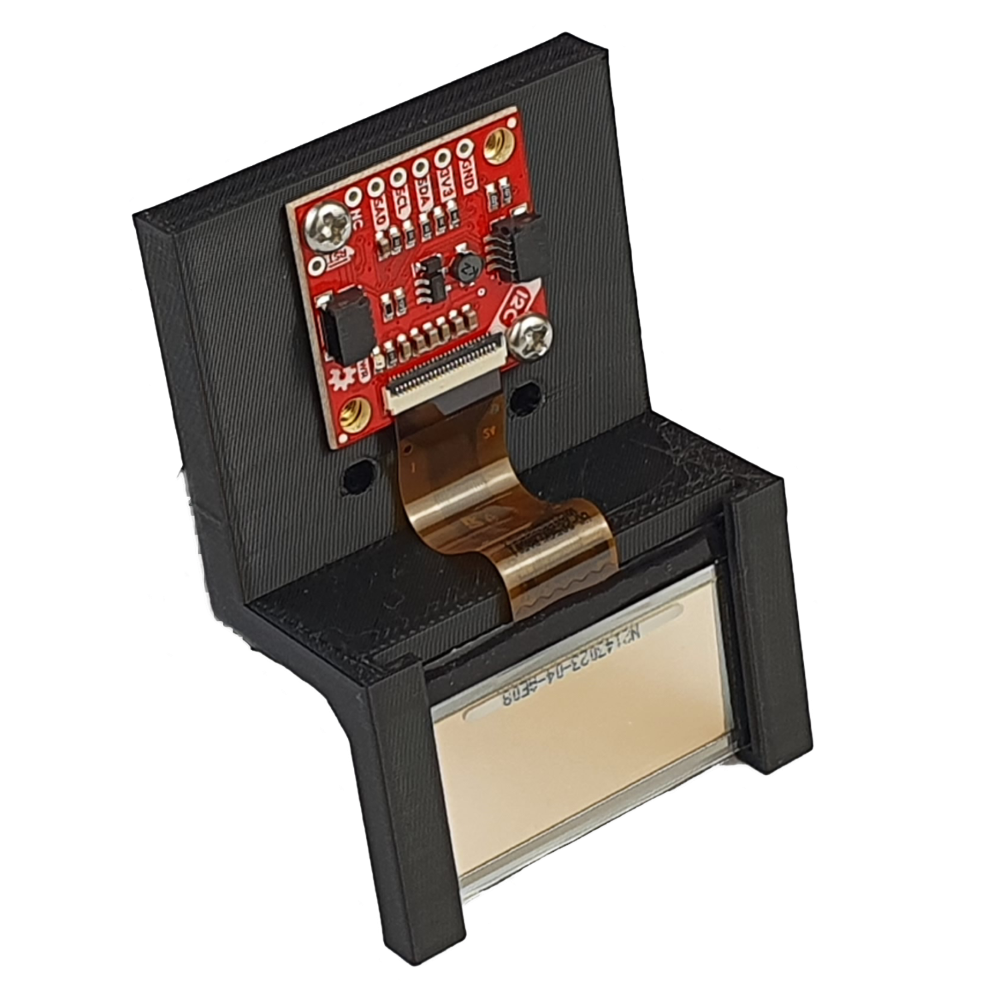}
        \caption{Transparent OLED display}
        \label{fig:module-oled}
    \end{subfigure}
    \hfill
    \begin{subfigure}[t]{0.19\linewidth}
        \centering
        \includegraphics[width=\linewidth]{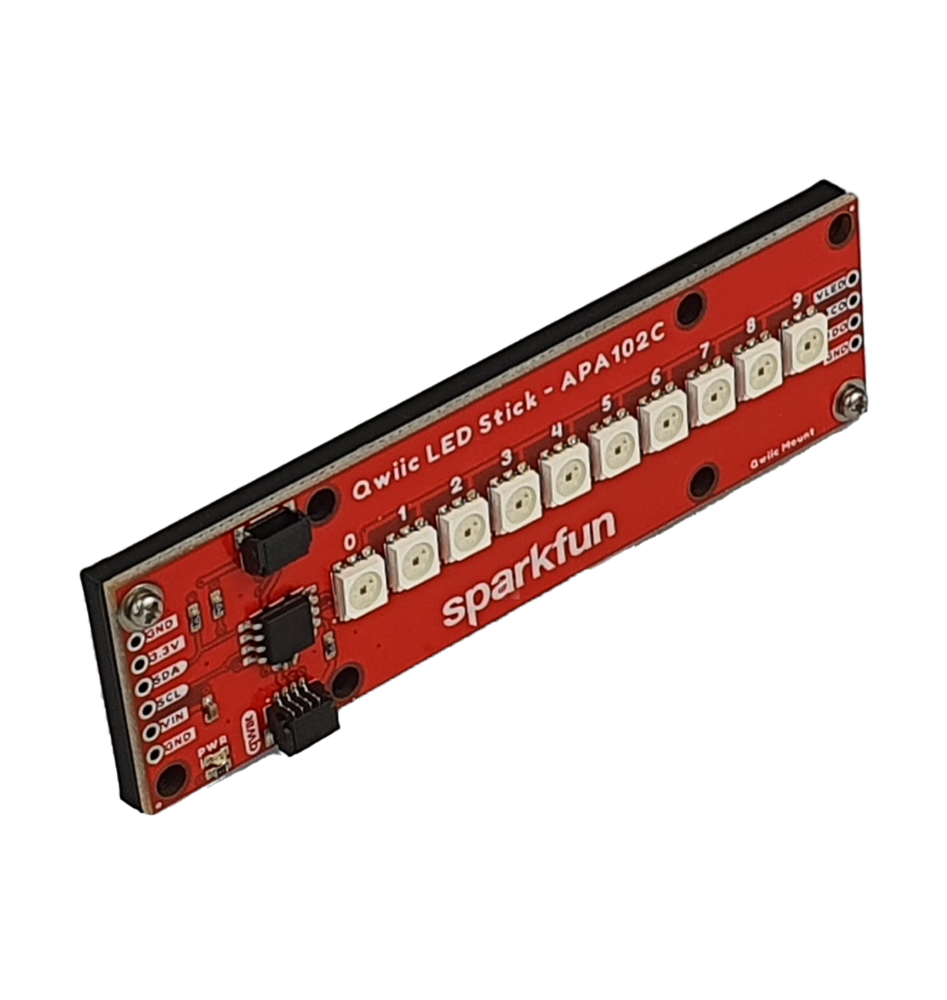}
        \caption{LED stick}
        \label{fig:module-led-stick}
    \end{subfigure}
    \hfill
    \begin{subfigure}[t]{0.19\linewidth}
        \centering
        \includegraphics[width=\linewidth]{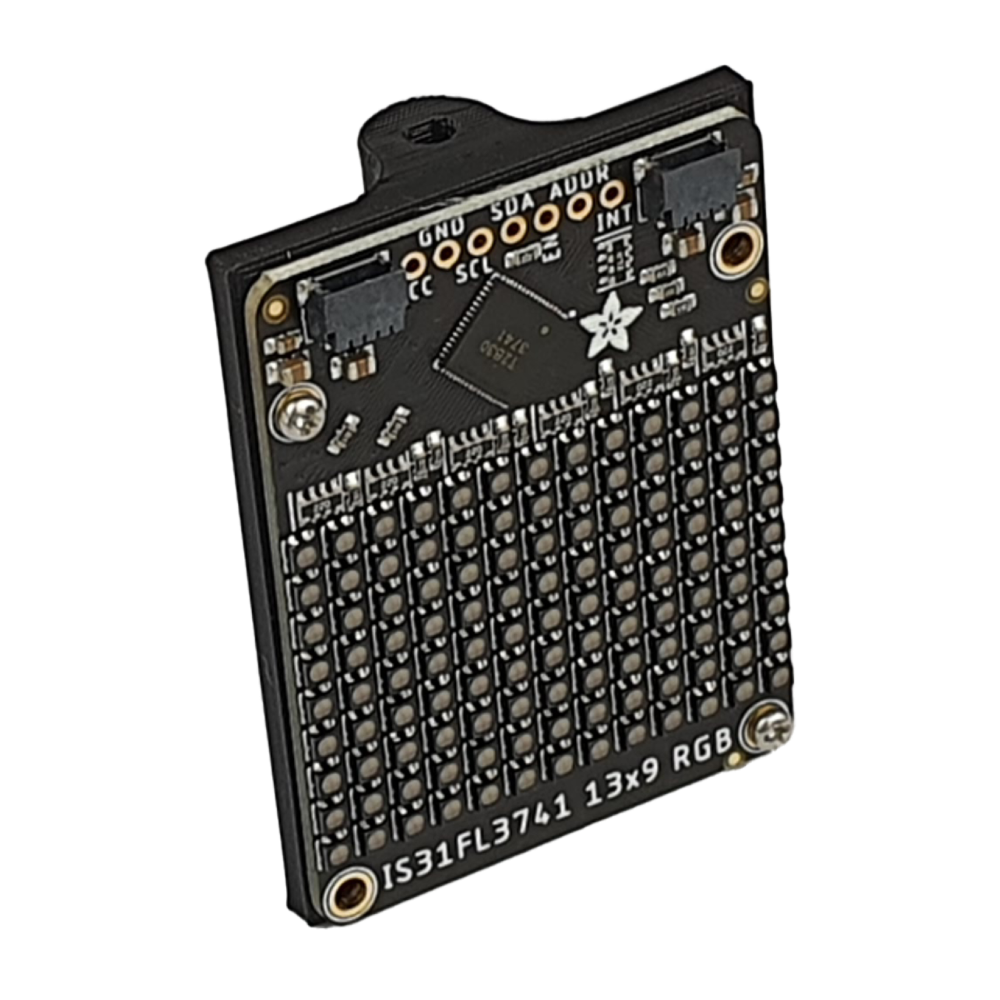}
        \caption{LED matrix}
        \label{fig:module-led-matrix}
    \end{subfigure}
    \vskip\baselineskip
    \begin{subfigure}[t]{0.19\linewidth}
        \centering
        \includegraphics[width=\linewidth]{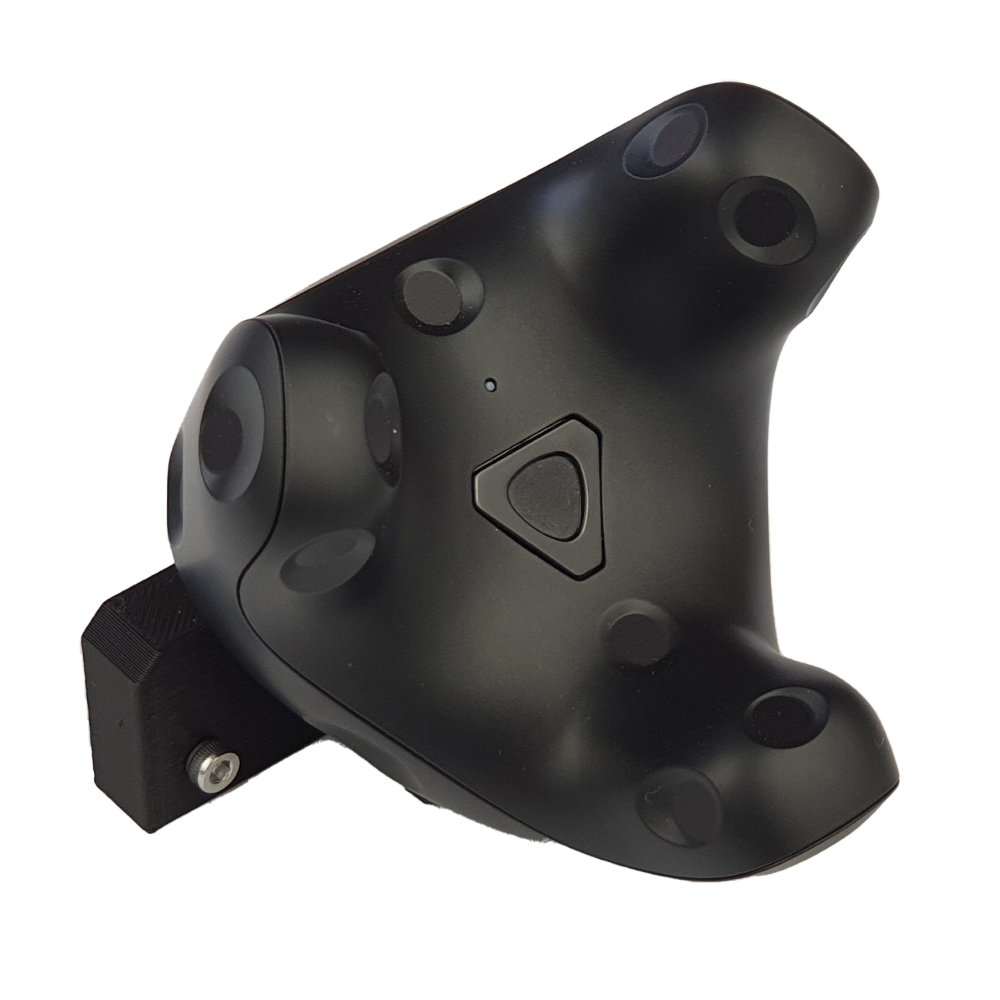}
        \caption{HTC Vive tracker}
        \label{fig:module-tracker}
    \end{subfigure}
    \hspace{1em}
    \begin{subfigure}[t]{0.19\linewidth}
        \centering
        \includegraphics[width=\linewidth]{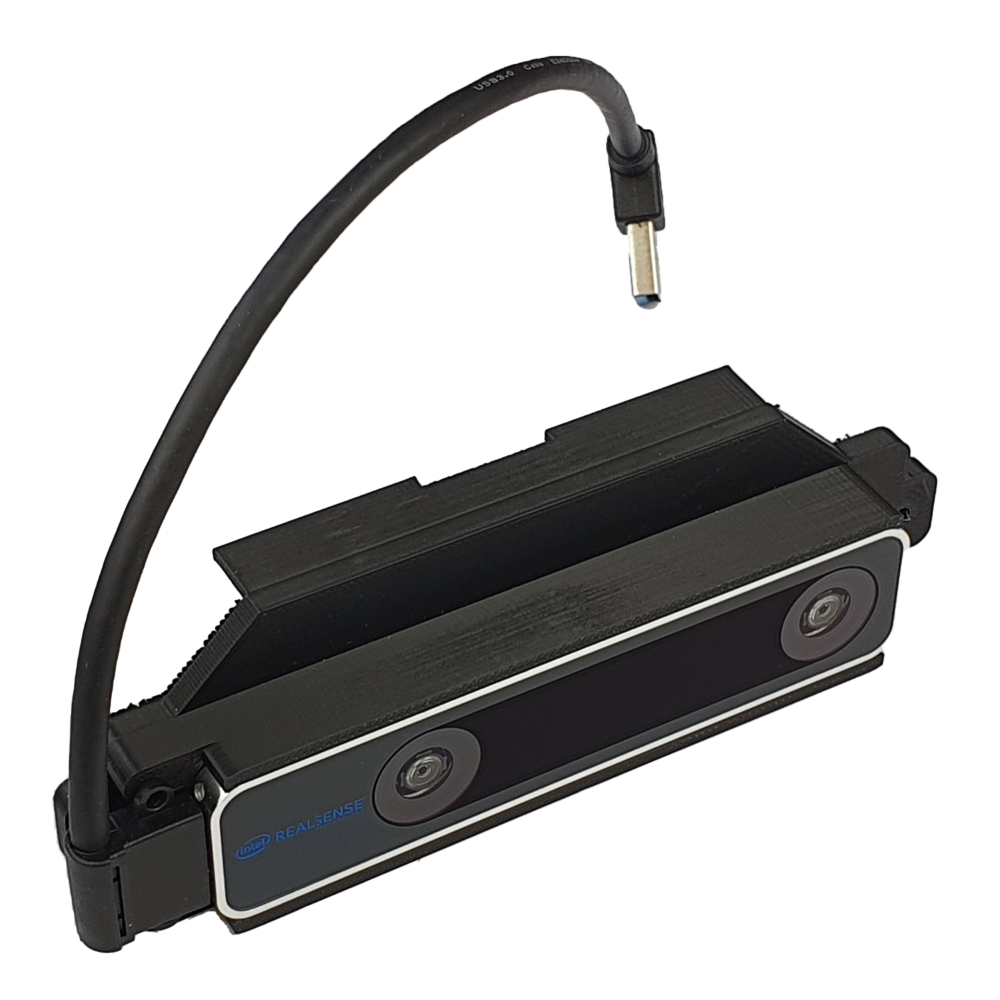}
        \caption{Intel T265 tracking camera}
        \label{fig:module-intel}
    \end{subfigure}
    \hspace{1em}
    \begin{subfigure}[t]{0.19\linewidth}
        \centering
        \includegraphics[width=\linewidth]{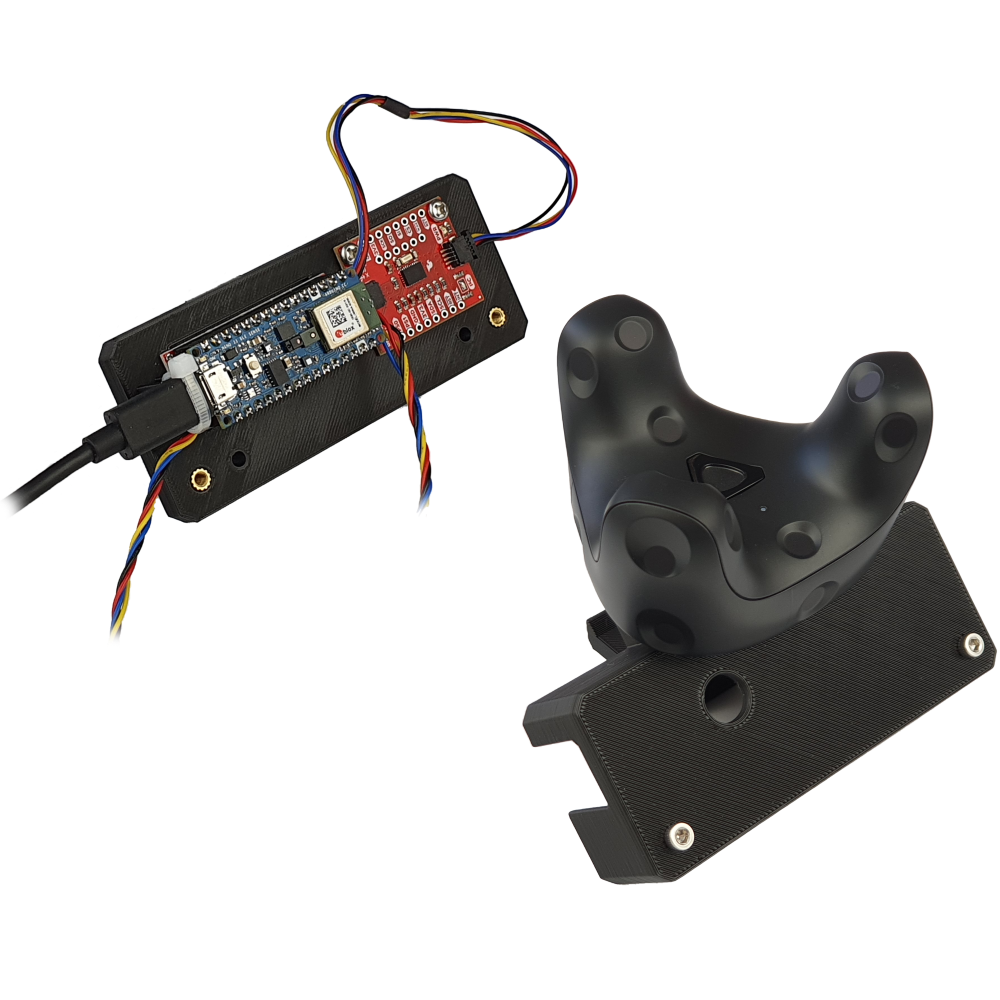}
        \caption{Arduino with IMU and optional tracker}
        \label{fig:module-arduino}
    \end{subfigure}
    \caption{Display and tracking modules.}
    \label{fig:display-modules}
\end{figure}

\begin{figure}
    \centering
    \includegraphics[width=\linewidth]{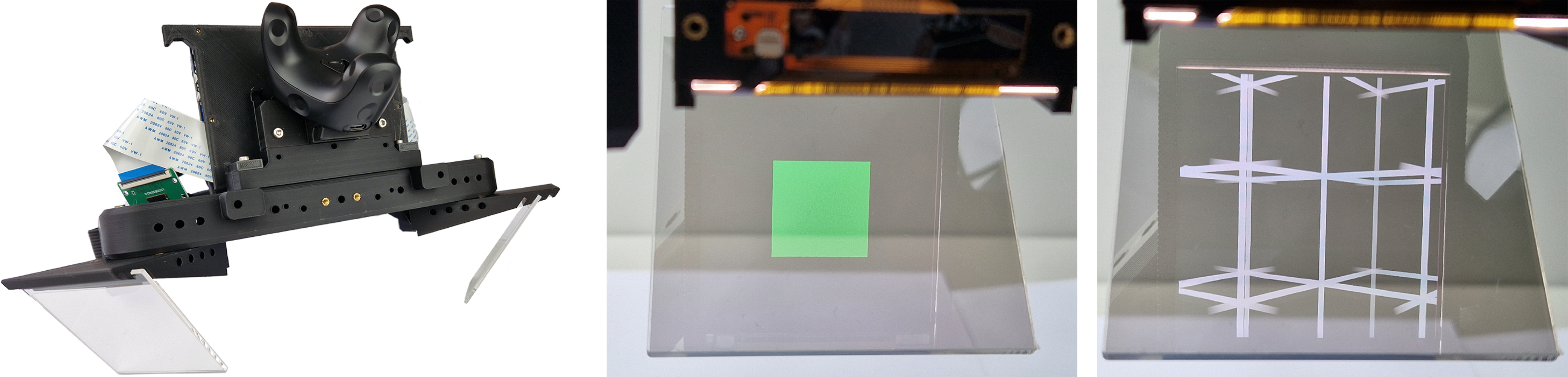}
    \caption{Configuration with see-through AR modules so the wearer can still see their surroundings.}
    \label{fig:configuration-ar}
\end{figure}

We use 3.1" displays (720x720 pixels, 60 Hz, JDI LT031MDZ4000) together with a converter board (Figure \ref{fig:module-lcd}). These regular LCD displays can be used to present a wide range of visual cues with respect to shape, size, fidelity, color, and brightness. 
Depending on the desired placement of the display modules, a notable amount of the user's peripheral field of view may be blocked. Therefore, we also feature see-through AR display modules that allow the user to perceive their real surroundings while still having peripheral cues displayed on top (Figure \ref{fig:configuration-ar}). For this module, we use 2.9" inch displays (1440x1440 pixels, 120 Hz, Sharp LS029B3SX02) that project onto a semi-reflective mirror foil glued on an angled acrylic glass. The foil is tinted to improve the contrast of the projected images in bright environments.

The LCD displays are driven via HDMI or DisplayPort for the video signal and USB for power. We chose these display models because they feature a high refresh rate, are reasonably small to attach near the eyes and are readily available on the market.
Using these displays enables plug-and-play. Because they act like normal monitors, one can easily show content on the displays without additional software requirements. 
The display modules offer mounting holes on each side and can either be screwed directly to the headset frame or with a rotatable spacer or linear rail to control the distance, height, and angle to the eyes (Figure \ref{fig:frame}). The display height was designed so that the center is approximately at the same level as the eyes, based on a model head and testing with users during development. 
Depending on a user's head shape and how they wear the headset, the exact location of the displays in relation to the head may vary. 
The purpose and focus of MoPeDT is peripheral vision. We did not design it to have display elements placed in the central field of view like regular VR or AR HMDs, but given the flexible mounting options, it is possible.

\begin{figure}
    \centering
    \includegraphics[width=\linewidth]{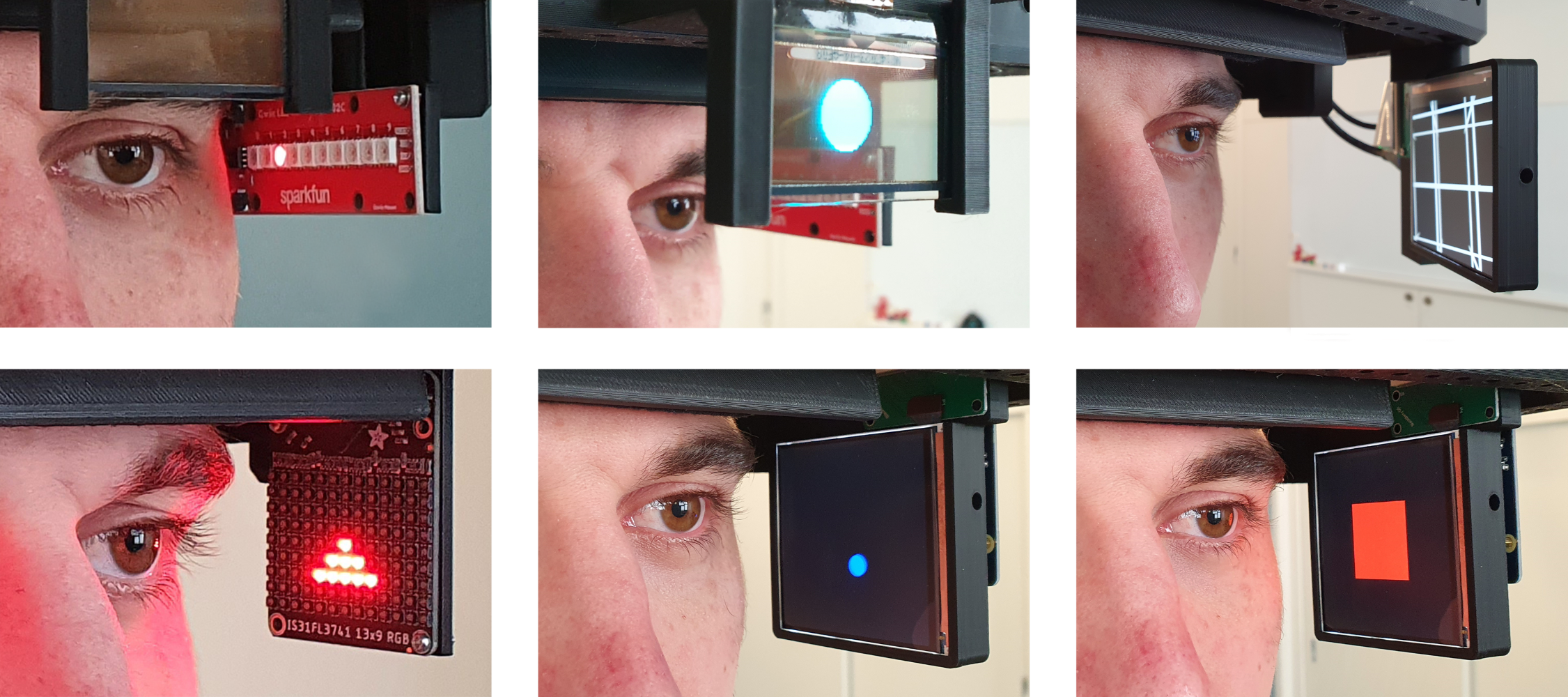}
    \caption{Overview of some peripheral visual cues on different near-eye displays.}
    \label{fig:mopedt-showcases}
\end{figure}

We also developed LED-based display modules. The LED displays are controlled using a compact Arduino Nano 33 BLE Sense microcontroller (Figure \ref{fig:module-arduino}) that features many on-board sensors that could be used in the visualizations. Additional sensors and electronics can be connected using SparkFun's Qwiic I$^2$C connection system \cite{qwiic} without soldering. This allows further modularization and rapid prototyping.
Rendering on these simple displays is much slower than using the LCDs due to bandwidth limitations of I$^2$C. Refreshing its content many times per second is not feasible when updating more than a few pixels/LEDs but is sufficient for simple animations.
The LED displays offer a less expressive visual language, due to their limited resolution and speed. Depending on the intended usage, they may be sufficient for some applications as peripheral detail perception is limited anyways \cite{strasburger_peripheral_2011}. They also have the advantage of being more lightweight, less bulky, and do not require a graphics card to drive them. The low power consumption of the Arduino makes it possible for MoPeDT to be battery-powered by a power bank and use it without carrying around a notebook.
We used a transparent graphical monochrome 128x64 OLED display (WiseChip UG-2856KLBAG01, 42 x 27.16 mm) from SparkFun (Figure \ref{fig:module-oled}). This module is also suitable for mounting in the near peripheral or more central field of view, as it allows see-through and does not occlude as much as the larger opaque screens. Similar to some related works \cite{nakao_smart_2016,gruenefeld_guiding_2018}, we also feature a SparkFun LED stick module (91.44 mm x 25.4 mm) and an Adafruit IS31FL3741 RGB LED 13x9 matrix module (51 mm x 39 mm) that can be used to show moving patterns of different colors (Figure \ref{fig:module-led-stick} and \ref{fig:module-led-matrix}). 
See Figure \ref{fig:configuration-ar} and \ref{fig:mopedt-showcases} for examples of different peripheral visual cues using MoPeDT.

For vision researchers, it is important to know the size, resolution, and location of a stimulus that is presented in the visual field. For MoPeDT, this can be easily be determined from the supplied CAD model by measuring the angle between the stimulus on a display module and the eyes of a model head from which the headset was designed (Section \ref{sec:frame}). We used Autodesk Fusion 360, in which the visual stimulus can be imported as a bitmap and positioned on the display modules. Angles and distances can then be obtained using the built-in \emph{Measure} tool. For individual user results, we used calipers to measure the relative eye positions to the headset and used them for the CAD measurements. Depending on the display module and the head of the user, we are able to display cues at up to 125 degrees horizontally on each side, exceeding the complete horizontal human field of view without eye movements \cite{strasburger_seven_2020}. Around 95 degrees of the vertical visual field can be used. However, the upper region is more constrained due to the head-worn frame.

\subsubsection{Tracking modules}

We provide three tracking solutions for the headset. The first uses SteamVR tracking via an HTC Vive tracker mounted above the front part of the frame. It is placed at an angle of 45 degrees (Figure \ref{fig:module-tracker} and \ref{fig:module-arduino}). We chose SteamVR lighthouse tracking due to its affordable cost, precision, and its widespread use in AR/VR labs.
Using the HTC Vive tracker works well in a laboratory environment but requires installed infrared lighthouses. 

We also implemented two tracking solutions that can be used outside and without the need for any other tracking equipment like the lighthouse stations. This makes it possible to use MoPeDT in real-life scenarios such as walking around in the city.
One available option is the Intel RealSense T265 tracking camera (Figure \ref{fig:module-intel}) which uses simultaneous localization and mapping to provide 3D poses in environments with a brightness of 15 lux or more \cite{intelt265}. This highly-integrated and affordable device in a lightweight package makes it an ideal fit for MoPeDT.
A third option is to use IMU-based rotation and acceleration with the Arduino module (Figure \ref{fig:module-arduino}). The module also comes with mounting holes for an additional Bosch BNO080 IMU breakout board, which offers higher accuracy and integrated sensor fusion compared to the on-board STM LSM9DS1 chip. This tracking solution is limited to 3D orientation and does not include translation. Although, it can work completely on its own without the need for a host computer to provide tracking data. 

All three tracking options can be used at the same time, for example, when using the Arduino for other sensor-related tasks or the Intel RealSense T265 as a camera alone. MoPeDT can also be adapted for motion-capturing systems like OptiTrack or Vicon by attaching active or passive markers to the frame.

\subsubsection{Software components}
As a starting point to develop applications for MoPeDT and for easy integration of existing AR/VR projects, we provide a Unity 2020 project with various presets and scenes. We provide Unity prefabs and C\# scripts for the camera setup to render for the peripheral displays with the correct rendering parameters already set up, ready-to-use peripheral visualizations, demo applications, and integrations of SteamVR, Intel RealSense, and Arduino-based tracking. To reduce latency for continuous translation (e.g. walking), we implemented pose prediction for the SteamVR and RealSense tracking options.
When using the LCD-based display modules, rendering is performed on the host computer (e.g. notebook in a backpack) to which the displays are directly connected.
For the Arduino side, we created several C/C++ sketches to control the LED and OLED modules.
The Arduino and attached display or tracking modules can be controlled from the host computer via a USB serial connection at 115200 baud by sending simple text commands from the Unity side. This also allows the Arduino side to use tracking information from SteamVR or Intel RealSense in the rendering that is provided by Unity.
With these software components, we give researchers a quick start after they have assembled the headset so they can directly start using it.
Following the idea of open science, the entirety of the software, CAD files for 3D printing, lists of necessary hardware components, and instructions can be downloaded and (re)used free of charge under the MIT license \cite{Albrecht_MoPeDT_2023}.  

\section{Application Use Cases}\label{sec:use-cases}
We envision multiple real-world scenarios where peripheral vision could be highly beneficial, leaving one's central field of view unobstructed. In the following, we motivate four very promising real-life applications of peripheral-only visual cues that take advantage of the perceptual limitations of peripheral perception. Previous research shows that simple peripheral out-of-focus cues can be understood by the user \cite{poppinga_ambiglasses_2012,luyten_hidden_2016,nakao_smart_2016}. We demonstrate how MoPeDT supports the development of peripheral vision applications by implementing prototypes for each of these use cases, following a validation by demonstration approach \cite{ledo_evaluation_2018}. The applications are a combination of existing peripheral visualization techniques, reimplemented using MoPeDT to show its versatility, or novel ideas on how peripheral vision could be utilized. All shown use cases are part of the MoPeDT repository and are freely available \cite{Albrecht_MoPeDT_2023}. 

\subsection{Enhancing spatial awareness with peripheral cues}\label{sec:spatial-awareness}

\begin{figure}
    \centering
    \includegraphics[width=\linewidth]{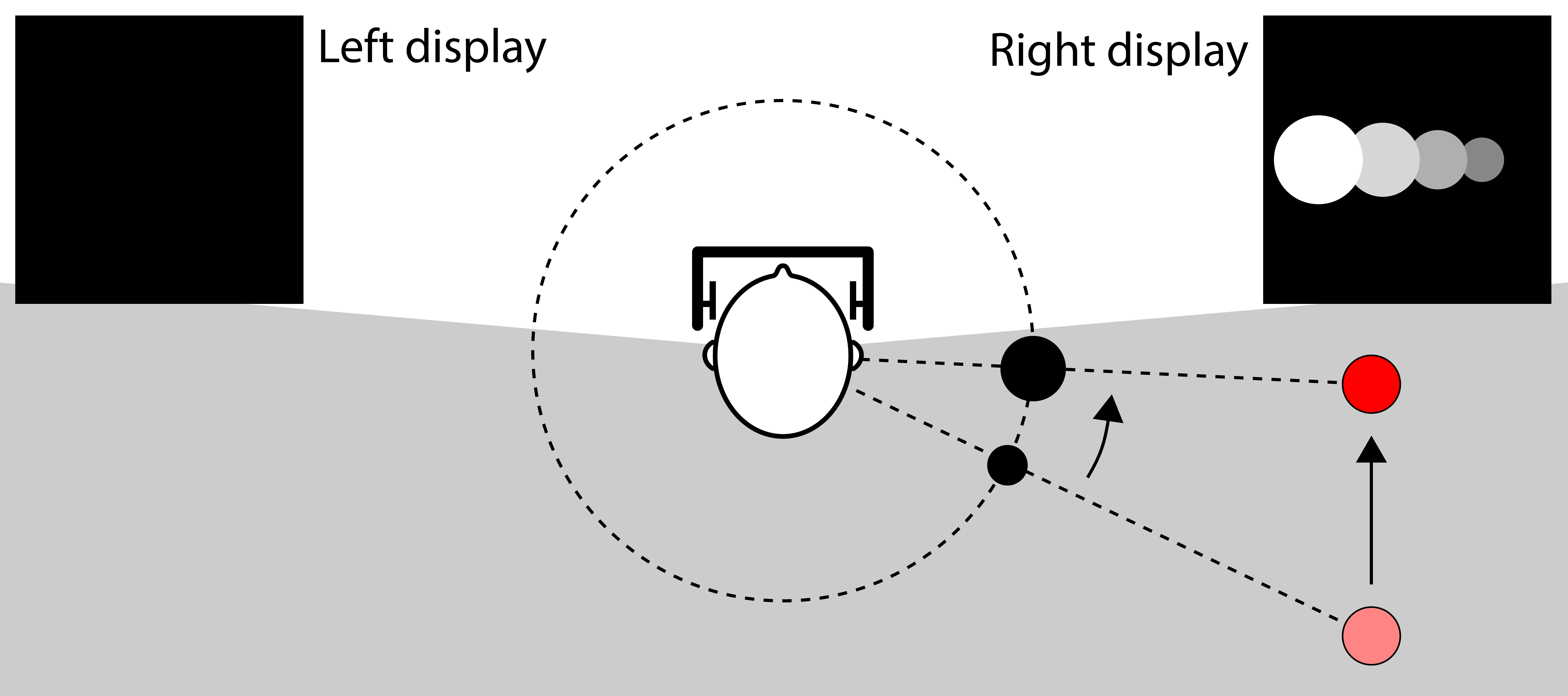}
    \caption{Top-down view of peripheral out-of-view visualization. A movement of an object in the rear back of the user (depicted as red circles) leads to a change of position and scale of the rendered sphere around the user's head (depicted as black circles), which can be seen on the right display. The gray area depicts the rendered field of view of the cameras, not of the user.}
    \label{fig:MoPeDT-oov}
\end{figure}

Spatial awareness is a prerequisite for orientation in and interaction with the environment. It affects personal safety, for example, detecting fast-approaching cars in traffic, and communicating with other people. Humans have a good awareness of objects inside their field of view. However, humans are often unaware of events outside their visual field. 
For traditional AR and VR, different out-of-view visualization methods have been developed \cite{gruenefeld_eyesee360_2017,gruenefeld_beyond_2018,gruenefeld_eyeseex_2018}.
Furthermore, spatial hearing enables us to locate sound sources with great precision, even when we are not able to see the object where the sound is coming from. Sound as an aid to locating out-of-view objects has been studied \cite{schoop_hindsight_2018,binetti_using_2021}. However, auditory cues are limited to events that make a distinguishable sound and are unavailable to deaf or hearing-impaired individuals who cannot rely on this natural mechanism. 

We propose using peripheral vision as an intuitive way to visualize and locate out-of-view objects by extending the natural human field of view beyond its limits. We make use of naturally existing visual event and object detection mechanisms.
Possible applications are being alerted of dangers behind the wearer like in traffic or when working with harmful machinery, augmenting spatial awareness for deaf or hearing-impaired persons who cannot rely on binaural hearing, navigation, and collaborating with others in either real or virtual environments. It might also prove useful for patients with central vision loss, such as age-related macular degeneration \cite{lim_age-related_2012}, which largely does not affect peripheral vision so prominent and moving shapes can still be recognized in the periphery.
To demonstrate this concept, we use MoPeDT in the configuration with parallel LCDs and an HTC Vive tracker (Figure \ref{fig:teaser-configuration-parallel}). The out-of-view object of interest is represented by an additional tracker. This visualization object (proxy) is rendered to the peripheral displays by two perspective rendering cameras that are placed at the user's tracked head position, essentially compressing the visual scene but preserving relative direction and speed of motions. As the periphery is especially good in detecting motion \cite{luyten_hidden_2016, palmer_vision_1999}, one can possibly relate the trajectory of the proxy on the displays to the movement of the actual physical object behind them. This takes advantage of the natural instinct to turn towards an object when one notices movement in the periphery. We assume that this feedback mechanism does not have to be learned or trained as it already happens subconsciously in everyday life. The proxy is rendered white on black to ensure good contrast, which is important for recognition in the periphery \cite{luyten_hidden_2016} (Figure \ref{fig:MoPeDT-oov}, top-down depiction).

We adapted a simplified version for the Arduino-controlled LED modules (Figure \ref{fig:module-arduino}). The transparent graphical OLED display module (Figure \ref{fig:module-oled}) is mounted in front, being visible in the upper part of the wearer's field of view. Two LED stick modules are attached to the side in the far periphery (Figure \ref{fig:teaser-configuration-led-stick}). This is conceptually very similar to some peripheral guidance and warning headsets \cite{gruenefeld_guiding_2018,tseng_lead_2015,niforatos_augmenting_2017}. The display elements show a moving light dot, thus prompting the user to turn left or right to the object of interest (Figure \ref{fig:mopedt-showcases}, first two images in the top row). By replicating the technical implementation by Gruenefeld et al. \cite{gruenefeld_guiding_2018}, we show how MoPeDT can be adapted for spatial guidance.

\subsection{Maintaining balance using peripheral vision}\label{sec:balance}
Vision improves standing and walking balance. Peripheral visual information contributes significantly to maintaining body posture \cite{paulus_visual_1984,nougier_contribution_1997,berencsi_functional_2005}. 
Balance impairments caused by disease or aging often lead to an increased dependence on visual cues \cite{lord_visual_1990}. This can result in falls when viewing a poor visual environment, such as when looking at a moving bus or a low contrasting scene. For elderly people, falls can lead to serious injuries and a general reduction in quality of life as they become scared of walking which only accelerates their increasing loss of mobility \cite{dionyssiotis_analyzing_2012}. 
Some authors have explored the usage of AR to provide balance and body posture training for the elderly \cite{kouris_holobalance_2018,mostajeran_augmented_2020}. Instead of using AR to improve balance in training sessions, we suggest a fundamentally different approach. 
We propose MoPeDT as an assistive device to improve balance by providing continuous peripheral visual feedback that is always available in everyday life, even in poor visual conditions. These peripheral cues could improve balance and reduce the risk of falling.

We render a grid of infinite horizontal and vertical white bars that are stable in relation to the surrounding environment and display them on the peripheral displays (Figure \ref{fig:mopedt-showcases}, top right). The orientation of the cues always stays relative to the direction of gravity. Tilting the head forward leads to a visual tilt backward of the bars on the displays, exactly like one would assume from real visual cues such as the outlines of a door or a cabinet. This way, the virtual cues behave like the real world and the headset user presumably does not have to learn the feedback mechanism. 
Again, the visual language we use here is simple, maximizing contrast and taking advantage of peripheral motion. The cues are even visible when the real environment does not supply rich enough visual information. The grid provides continuous feedback about the wearer's posture.
For this demonstration, we use a different configuration of MoPeDT (Figure \ref{fig:teaser-configuration-angled}). Based on a model head, the centers of the LCDs are placed at an angle of 52 degrees eccentricity using the normal frame so the cues are more prominent compared to having them parallel to the side of the head. Each display covers 35 degrees of the total visual field. The thickness of a drawn bar roughly covers 0.38 - 0.51 degrees of visual angle at around 24 pixels per degree. 
Improving balance using peripheral cues is a scenario that should also be possible outside the lab, where no permanently installed tracking solutions are available. Hence, we use the Intel RealSense T265 tracking camera module here. 
A notebook carried in a backpack can be used to perform the rendering as the headset is tethered in this configuration. This way, the entire device is self-contained and can be used while walking outside.
Using peripheral vision for balance improvement showcases how MoPeDT could also be used beyond the usual scope of AR and HCI for interdisciplinary research that benefits from mutual findings in either field.

\subsection{Peripheral interaction}\label{sec:peripheral-interaction}
In some scenarios, being distracted could lead to potentially life-threatening situations, for example, using a smartphone while driving. Peripheral user interfaces do not clutter and conceal central vision like many conventional user interfaces in HMDs. To demonstrate peripheral vision for interaction, we propose a novel selection method using handheld VR controllers. Our technique does not require a gaze shift towards the user interface and also relies on proprioceptive and haptic cues so the user can stay focused on the task at hand. We show how peripheral cues may also be used to actively manipulate something instead of only receiving passive information about something happening. Luyten et al. found that interactive tasks on peripheral near-eye displays can be accomplished with great precision \cite{luyten_hidden_2016}. 
Our method is closely inspired by eyes-free target acquisition techniques that leverage spatial memory and proprioception to allow a user to acquire an object in virtual space without directly looking at it \cite{cockburn_air_2011,yan_eyes-free_2018,zhou_eyes-free_2020,wu_exploring_2021}. However, unlike these approaches, we use peripheral feedback cues to visualize the selection. Our technique is similar to hand movement and grasping using peripheral vision in real life. For example, grabbing a glass of water is possible without directly looking at it.

\begin{figure}
    \centering
    \includegraphics[width=\linewidth]{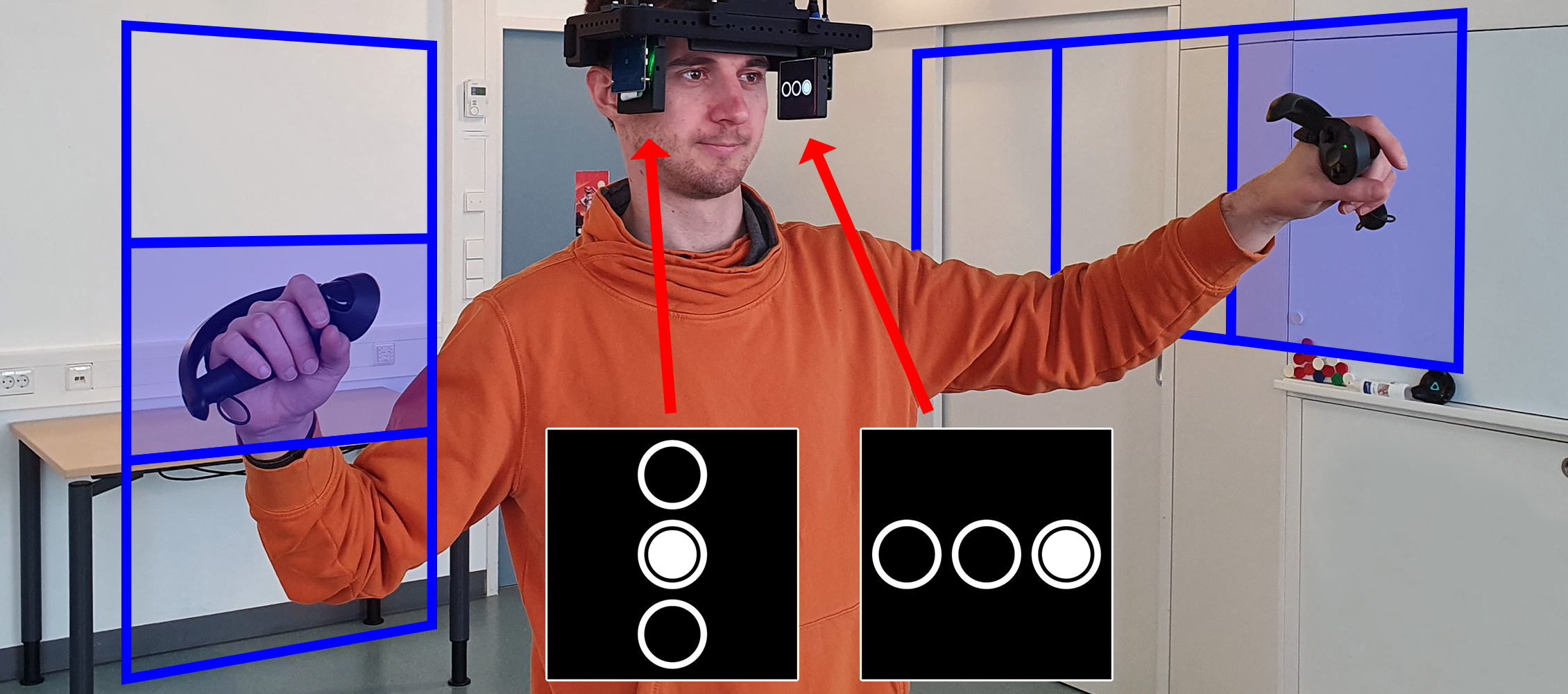}
    \caption{Illustration of volumes used for peripheral interaction. For improved clarity, we show a 2D representation of the 3D bounding boxes. The corresponding selected options are shown on the near-eye displays as filled circles.}
    \label{fig:mopedt-interaction}
\end{figure}

For this showcase, we employ the configuration of MoPeDT with parallel LCDs and an HTC Vive tracker (Figure \ref{fig:teaser-configuration-parallel}). In addition, two Valve Index controllers are used for selection.
With each hand, the user can select three different options via hand and arm movement (Figure \ref{fig:mopedt-interaction}). On both sides of the head, three adjacent bounding boxes are placed within arm's reach. Depending on which 3D volume the tracked controller currently is inside, its corresponding option is selected. The options are shown on the appropriate peripheral display by three circles that are filled when they are selected. Additionally, a haptic pulse on the corresponding controller is fired when an option is selected or changed. 

\subsection{Self-adaptation through peripheral visual feedback}\label{sec:self-adaptation}
Adapting one's behavior based on feedback from others or the environment is a common everyday task. An application for real-time feedback during public speaking using peripheral vision was proposed by Luyten et al. \cite{luyten_hidden_2016}. Using their developed guidelines for peripheral cues, they suggested visualizing pace, volume, and energy by three moving shapes in different colors to improve the performance of the speaker. As another example of our implementation of this concept, we envision an athlete running and monitoring their heart rate. To maintain a constant training load, it should stay at a certain level (e.g., 150 BPM), not falling below and not exceeding it. 
We do not use actual heart rate data in this demo and instead simulate the measured value by an on-screen slider. 
A deviation of the target heart rate is visualized by moving arrows (Figure \ref{fig:mopedt-showcases}, bottom left). 
For this showcase, we employ the LED matrix configuration of the headset (Figure \ref{fig:teaser-configuration-led-matrix}). This showcase demonstrates how we can employ MoPeDT to implement peripheral HMD approaches that use LED matrices with animated cues \cite{nakao_smart_2016,gruenefeld_guiding_2018}.
Again, our implementation leverages a simple visual language, relying on peripheral motion perception. While arbitrary, the direction of the arrow and its motion follow an intuitive understanding that up is associated with a high value and vice versa.  
Compared to a smartwatch, peripheral feedback is continuously available and the user does not have to look at their wrist. Unlike audio cues, it is available for deaf or hearing-impaired people.

\subsection{Peripheral notifications}\label{sec:notifications}
Notifying a user with subtle peripheral visual cues is a generalization of the concepts and techniques described above. In all previously described scenarios, a change is visualized, prompting or reminding the wearer to perform some kind of task, either consciously or subconsciously. Costanza et al. first explored personal and intimate peripheral notifications in eyeglasses and found that they are noticeable while doing everyday tasks \cite{costanza_eye-q_2006}. Peripheral notifications could be used as a more subtle method of being notified of messages, reminders, and similar. Compared to vibration and sound, they are silent and do not alert others. Visually they are potentially less disruptive than displaying content in central view. 
As a final demonstration of the flexibility of MoPeDT and how it can aid peripheral vision research, we show how it can be used for general-purpose notifications by replicating and extending technical implementations presented in previous works \cite{costanza_eye-q_2006, luyten_hidden_2016}. Using the LCD or LED modules, our application allows the selection of multiple shapes (circle, filled circle, square, triangle) in different colors (red, green, blue, white) with optional animations (moving up/down/right/left, rotating, growing and shrinking, popping up from a side, flashing, blinking) at different sizes and speeds (Figure \ref{fig:mopedt-showcases}). As mentioned before and in line with the other showcases, the visual language is simple, taking advantage of the unique perception properties of peripheral vision and relying on developed guidelines for such cues \cite{palmer_vision_1999,strasburger_peripheral_2011,luyten_hidden_2016}. 
Beyond the shapes and animations discussed here, many of the other visual elements we showed so far (Figure \ref{fig:mopedt-showcases}) could be used for peripheral notifications, either using LCD or LED displays. Our toolkit allows exploring this design space.

\section{Evaluation}
We already explored sample applications of peripheral-only cues in detail in Section \ref{sec:use-cases}. We also collected initial usability metrics about MoPeDT as part of an evaluation of the balance-improving potential of peripheral cues (Section \ref{sec:user-study}). In addition to demonstration \cite{ledo_evaluation_2018}, we performed a heuristics-based evaluation \cite{olsen_evaluating_2007} (Section \ref{sec:heuristic-evaluation}). Some evaluation methods have been proposed to evaluate ambient and peripheral displays \cite{matthews_defining_2007,mankoff_heuristic_2003}, but those were not developed with HMDs in mind and thus were unsuitable for our system.

\subsection{User Study}\label{sec:user-study}
We ran a user study to check initial user acceptance and test if MoPeDT is suitable for further experiments that involve users wearing the headset. The user study was initially designed in order to test if peripheral visual cues improve balance during standing. The physiological balance results are out-of-scope for this paper and will be subject to other publications. Here, we concentrate on the subjective feedback.

A total of 15 participants (8 female, 7 male) between ages 20 and 35 (M = 24.60, SD = 3.72) took part in 10 balance-related measurements, each 1 minute long with short breaks while wearing MoPeDT and having balance cues displayed. Afterward, they were handed a questionnaire asking about subjective comfort and experience while wearing the headset. See Section \ref{sec:balance} for details about the visual stimuli and the used hardware configuration of MoPeDT in this experiment. The study was conducted in agreement with the IRB regulations by the ethics committee of the University of Konstanz and the latest version of the Declaration of Helsinki. 
Answers were given on a 5-point Likert-scale (disagree 1 to agree 5):

\begin{enumerate}
    \item I was irritated by the cues on the displays in the peripheral field of view. (M = 2.27, SD = 0.88)
    \item I found the headset comfortable to wear. (M = 3.47, SD = 1.41)
    \item I found that my field of view was reduced while wearing the headset. (M = 3.40, SD = 1.24)
    \item I found the headset annoying. (M = 2.20, SD = 1.08)
\end{enumerate}

\begin{figure}
    \centering
    \includegraphics[width=\linewidth]{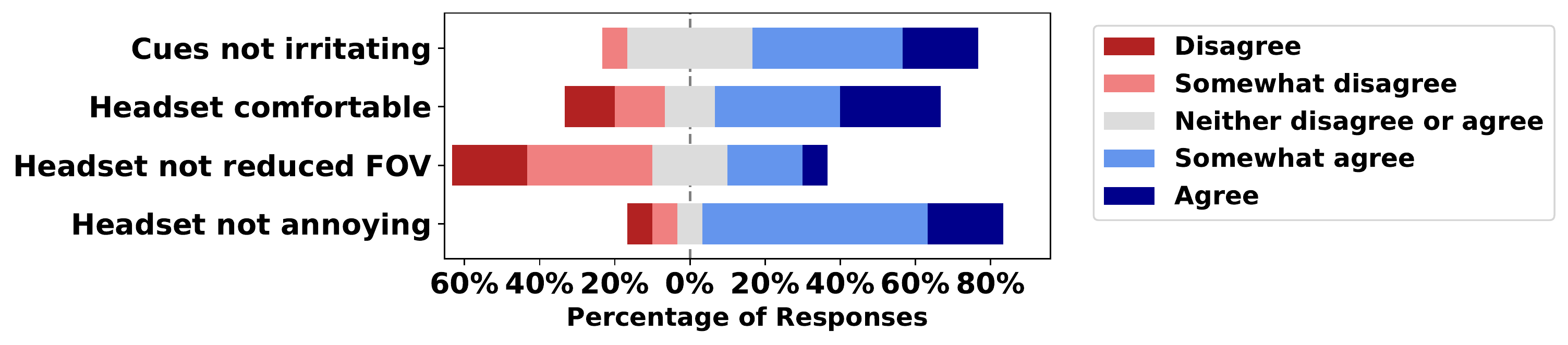}
    \caption{Answers to Likert-scale questions about subjective user experience while wearing MoPeDT. Items 1, 3, and 4 are reversed.}
    \label{fig:subjective-likert}
\end{figure}

See Figure \ref{fig:subjective-likert} for the distribution of answers. Reversing items 1, 3, and 4 for coherent sub-scales gives an average total score of 13.60 out of 20 (SD = 2.69). A higher score indicates a better overall subjective experience. We also asked participants to report if they felt dizzy at any point while wearing the headset. One individual stated that they had minor headaches from the pressure of the headset on the head. Three persons reported a slight unspecified feeling of dizziness.

\subsection{Heuristic Evaluation}\label{sec:heuristic-evaluation}
In the following, we, the authors themselves, evaluate MoPeDT based on the heuristics for user interface systems proposed by Olsen \cite{olsen_evaluating_2007}. We follow a heuristic checklist approach \cite{ledo_evaluation_2018}.

\paragraph{Situations, Tasks, and Users (STU)}
MoPeDT is a hardware and software toolkit to prototype, iterate, and study (Task) peripheral vision applications for everyday scenarios (Situations). It is primarily used by researchers and developers (Users). 

\paragraph{Importance}
Having an adaptive and flexible development platform like MoPeDT is important to create peripheral vision prototypes, quickly try out various configurations, and evaluate those in user studies. We showed in Section \ref{sec:use-cases} that there are compelling applications of peripheral vision that justify the need for a specialized toolkit that previously did not exist.

\paragraph{Problem not previously solved}
Until recently, researchers interested in studying peripheral-only cues had to come up with their own hardware design as they could not rely on a ready-to-use solution as they are available for traditional AR and VR (see Section \ref{sec:related-work}). This poses a significant entry barrier for further research. With MoPeDT, we provide the first multi-purpose HMD toolkit specifically tailored for peripheral vision applications that allows quick reconfiguration. 

\paragraph{Generality}
To show the generality of MoPeDT, we demonstrated different configurations with different modules (Figure \ref{fig:teaser-configurations}) and showcased them in distinct scenarios, each of them making use of different display modalities and tracking requirements. The design of the head-mounted frame and modules is intentionally simple yet versatile with its many mounting points so MoPeDT may adapt to use cases it was not specifically designed for. The many available sensor and display modules (Figure \ref{fig:display-modules}) allow a selection of desired visual fidelity, tracking requirements, environmental sensing, and weight.

\paragraph{Reduce solution viscosity}
MoPeDT supports \emph{flexibility} with a number of modules that can be quickly attached or removed during an exploratory design phase or even while conducting an experiment. Instead of a fixed arrangement of components, they can be exchanged, the position and eccentricity of peripheral cues can be finely controlled, and the whole headset can be reconfigured. 
For example, Gruenefeld et al. decided to switch from an LED strip to an LED matrix based on expert user feedback during development \cite{gruenefeld_guiding_2018}. This required designing a new pair of glasses. With MoPeDT, such a desired module is already available.
With a set of reusable modules and pre-defined visualization components for peripheral cues, we achieve \emph{expressive leverage}. They are a unification of previous approaches of peripheral HMDs (e.g., LED displays \cite{costanza_eye-q_2006,gruenefeld_guiding_2018}, LCD screens \cite{nakano_head-mounted_2021, luyten_hidden_2016}) or based on confirmed findings about peripheral visual perception \cite{luyten_hidden_2016}. This way, new users of MoPeDT can start with a toolkit of proven concepts and do not have to make these choices themselves.
Peripheral visualization techniques created with MoPeDT make use of the natural real-world perception in the periphery by facilitating simple shapes, strong contrasts, and motion \cite{palmer_vision_1999,strasburger_peripheral_2011,luyten_hidden_2016}, thereby increasing its \emph{expressive match}. 

\paragraph{Empowering new design participants}
The LCD modules of MoPeDT work like a regular monitor. Users do not need to be experienced with technology or programming to create visualizations. For simple experiments, playing a video with suitable peripheral cues is sufficient. Previous peripheral HMDs like \cite{costanza_eye-q_2006} or \cite{luyten_hidden_2016} require some technical expertise to get even simple visualizations working.
Researchers from a non-technical field may more easily contribute to the design and development process with their unique expertise. However, assembling the headset itself and more complex visualization techniques require some technical knowledge.

\paragraph{Power in combination}
To enable \emph{inductive combination}, MoPeDT provides a set of powerful building blocks that can be combined to create versatile configurations of the headset for different applications as shown above. 
We \emph{simplify interconnection} by using HDMI, DisplayPort, USB, and the solder-less Qwiic I$^2$C system, allowing easy integration and communication of components.
Through the use of standard frameworks (Unity, SteamVR, Arduino), pre-defined peripheral visualizations, standard connectors, off-the-shelf plug-and-play electronics, and simple serial communication protocols (Section \ref{sec:implementation}), we enable \emph{ease of combination}.

\paragraph{Can it scale up?}
The modularity of MoPeDT with its flexible hardware and software components may allow our system to scale to larger problems and demands. We demonstrated this by showing diverse showcases of the headset in different configurations. New custom modules, may be created using 3D printing and integrated with the software framework. This fosters extendability. Some particular or exotic peripheral vision applications might be beyond the physical or technical capabilities of our system, but we think that the offered components give a lot of creative freedom to explore the design space of peripheral-only cues.

\section{Discussion, Limitations and Future Work}

From our implemented peripheral vision prototype applications and the heuristic evaluation we conclude that MoPeDT is an effective and novel tool for prototyping and testing new applications using peripheral-only cues. The modularity of MoPeDT, the variety of available tracking and display modules, flexible mounting options, the integration with established software and hardware frameworks, read-to-use peripheral visualizations, and open-source rationale allows researchers to quickly build and iterate peripheral vision solutions without the need to develop their own specialized hardware and software. In addition, our selection of hardware and software components reduces the initial effort to decide which components are necessary and suitable for peripheral vision research. Our example visualizations make use of different configurations to explore what is possible with MoPeDT so researchers can decide which hardware and software configuration work best for their desired application. 
Thus, our system is more general and more versatile than related devices that were developed for specific applications of peripheral vision \cite{costanza_eye-q_2006,poppinga_ambiglasses_2012,tseng_lead_2015,nakao_smart_2016,luyten_hidden_2016,niforatos_augmenting_2017,cobus_multimodal_2017,van_veen_situation_2017,gruenefeld_guiding_2018,olwal_1d_2018,kiss_clairbuoyance_2019}.
Existing HMD toolkits were either limited to specific means of peripheral cues like radial LEDs \cite{gruenefeld_perimr_2017} or peripheral vision was not a key aspect of their system \cite{endo_modularhmd_2021}.
Due to technical differences in all those approaches, such as different display placement, display electronics, or dimensions, comparing them in a fair manner is challenging. 
Having access to a unified and multi-purpose toolkit like MoPeDT to conduct peripheral vision research allows for better comparability of the solutions that were created with it. 
The modular approach is one of the key advantages of MoPeDT. In addition, AR/VR and HCI researchers are most likely already familiar with the used software and hardware frameworks such as the Unity engine, SteamVR, and Arduino.
We showed how one can easily reimplement different peripheral guidance and notification techniques from previous works and try out new concepts using our system.

From implementing these peripheral vision applications using MoPeDT, we gained some insight into working with the toolkit as a whole. We often experimented with different spatial arrangements of the display elements and valued the ability to exchange components quickly. 
We found that at larger eccentricities, peripheral cues are less distracting, block less of the field of view, and can still easily be perceived. For most use cases, we suggest using a configuration with parallel display elements (Figure \ref{fig:teaser-configuration-parallel}, \ref{fig:teaser-configuration-led-stick}, and \ref{fig:teaser-configuration-led-matrix}) or using see-through display elements (Figure \ref{fig:configuration-ar}). Using Unity instead of creating our own renderer for the LCD displays proved especially helpful because the prefabs for MoPeDT allowed us to quickly create peripheral visualization scenes. They abstract the rendering process and help with the camera setup. 

Peripheral vision applications might prove useful in various domains of AR, VR, HCI, and beyond. An interesting application might be to apply peripheral cues to assist physically challenged users. For instance, deaf people's perception is limited to their natural field of view  (200 to 220 degrees horizontally \cite{strasburger_seven_2020}), while hearing people's auditory system provides a 360-degree perception of the environment through sound cues. Artificially enhancing deaf people's perception of their surroundings, e.g. about a car approaching from behind, by means of unobtrusive peripheral cues (Section \ref{sec:spatial-awareness}) might enhance their spatial awareness and improve their safety and quality of life. Balance improvement using peripheral vision (Section \ref{sec:balance}) for elderly people could drastically reduce the risk of injuries caused by falls and give people more confidence when walking. Balance cues in the periphery might also be used to mitigate motion sickness in VR as they give the user a constant reference of their body. Peripheral visual cues could also pose useful for individuals suffering from central vision loss (e.g. age-related macular degeneration \cite{lim_age-related_2012}). Consequently, it is crucial to give researchers the necessary tools to create and evaluate such peripheral visual aids more easily. Hence, our system may not only be relevant for researchers in AR/VR and HCI, e.g. to design future HMDs, but might be a useful research tool for other researchers from sports science, neuroscience, psychology, and cognitive science.   

The results from the user study suggest users were largely not irritated by the balance cues that were shown in their peripheral vision and the headset itself was not annoying. A subjective reduction of the field of view was expected as we used opaque display modules. At the time of the study, the see-through AR module was not implemented yet which would have likely alleviated this problem. We think that the subjective results show that MoPeDT is reasonably well received by users for its intended purpose of peripheral vision research and is suitable for further experiments involving users wearing the headset. Still, we have to find ways in future design iterations to prevent dizziness and improve wearing comfort to make MoPeDT more suitable for longer periods of usage.

Depending on the used modules and due to the variety of mounting options, the headset can become quite bulky and heavy.
Other peripheral glasses like \cite{costanza_eye-q_2006}, \cite{luyten_hidden_2016}, and \cite{gruenefeld_guiding_2018} are more compact and lightweight, also potentially more comfortable to wear. However, they do not feature the adaptability of MoPeDT. During our design process, we prioritized configuration flexibility over comfort, which is more important during a prototyping phase where things are not settled yet and requirements change. 
As MoPeDT was designed for peripheral-only augmentations, the size and shape of the present implementation of our headset make it difficult to wear it together with another large AR headset like the Microsoft HoloLens 2. More compact devices in the shape of regular glasses would fit. The peripheral cues could add additional information to the AR scene, such as the position and relative motion of bystanders (see Section \ref{sec:spatial-awareness}) in collaborative scenarios. Such a combination of central and peripheral visual augmentation would be very interesting to study in the future but is currently a limitation of our toolkit. 

Peripheral cues can evoke a weak accommodation response which is also influenced by eye convergence and rapidly declines towards 30 degrees eccentricity \cite{gu_accommodation_1987}. In our headset configurations, we tried to minimize binocular overlap by placing display modules at large eccentricities to avoid the vergence-accommodation conflict (VAC) \cite{hoffman_vergenceaccommodation_2008}. MoPeDT currently focuses primarily on simple out-of-focus peripheral-only cues in the far periphery ($>$ 45 degrees), where the accommodation response is especially weak. As such, we suspect it can be neglected for this kind of peripheral vision research as research shows that simple peripheral cues can be understood correctly by users \cite{poppinga_ambiglasses_2012,luyten_hidden_2016,nakao_smart_2016}. To present a sharp virtual image further away like conventional AR glasses, especially towards central vision, accommodation-supporting near-eye displays with more sophisticated optics may be necessary \cite{koulieris_near-eye_2019}.

A key concept of the envisioned peripheral augmentation we want to achieve with our device is that the central view stays unobstructed.
However, depending on which display modules are used and how they are attached to the frame, for example, the angled opaque 3.1" LCD screens (Figure \ref{fig:teaser-configuration-angled}), the field of view of the wearer is narrowed. The display elements act like blinders. For this reason, we developed the see-through AR modules (Figure \ref{fig:configuration-ar}). However, in bright environments, the peripheral cues are hard to see due to the relatively low brightness of the LCD displays when projecting onto the reflective glass, even with the tinted foil. More specialized AR micro-projectors or waveguide displays would be necessary to mitigate this, but those solutions are very expensive.
Nakano et al. increased the downward field of view with additional displays \cite{nakano_head-mounted_2021}. Such a placement of peripheral displays would be possible for MoPeDT with a special mounting adapter. Placing displays above eye level is difficult beyond a limited angle where the frame sits on top of the forehead. For such a configuration, MoPeDT might not be suitable or would require designing a completely new frame.

As of now, our system always requires a graphics card to drive the LCD screens or at least a USB power source for the Arduino-based display modules. This means that the headset is tethered. For usage outside the lab, a notebook in a backpack or mobile power supply is required. We think this is sufficient for prototyping, but a more integrated, standalone solution with electronics and batteries could give greater flexibility.

Some authors suggest that peripheral cues are only useful to a certain degree \cite{sun_investigating_2018,kishishita_analysing_2014}. At higher eccentricities, the perceptibility of peripheral information decreases and response time increases. Under high workloads or demanding tasks, peripheral vision can temporarily degrade \cite{costanza_eye-q_2006,williams_tunnel_1985}. 
Foveated rendering exploits the unique perceptual characteristics of peripheral vision to improve performance for real-time rendering in AR/VR \cite{patney_towards_2016,kaplanyan_deepfovea_2019}. Performance is no issue for MoPeDT as the rendered cues are simple without requiring expensive lighting or shading techniques. However, some of the vision foundations of foveated rendering should be kept in mind when designing peripheral visualizations. At larger eccentricities, stimuli size should be increased to give a similar perceptual appearance (cortical magnification) \cite{strasburger_peripheral_2011}.
As the periphery is sensible to flickering, temporal aliasing should be kept to a minimum \cite{patney_towards_2016}.
Distinguishing multiple objects in the periphery is challenging, also known as crowding \cite{levi_crowding_2009,luyten_hidden_2016}. This limits the number of cues that could be sensibly visualized and recognized. 

MoPeDT is actively used for sports science experiments. Those results will be subject to upcoming publications. In this paper, we wanted to concentrate on MoPeDT as a toolkit and show application areas where it could aid peripheral vision research. In the future, we want to evaluate specific use cases of peripheral visualization and interaction techniques that we showed in Section \ref{sec:use-cases} in more detailed user studies to investigate their effectiveness and benefit.

\section{Conclusion}
We introduced MoPeDT, a modular headset toolkit to research peripheral vision applications for AR/VR, HCI, and related fields using different near-eye displays and tracking options. 
We demonstrated the benefit and versatility of our system by implementing promising applications for peripheral cues.
Our user study suggested that MoPeDT is suitable for further experiments, but comfort could be further improved.
In addition, we performed a heuristic evaluation to highlight how MoPeDT could be applied to a general set of peripheral vision research questions. 
Furthermore, we critically reflected upon current limitations like its restricted combination with other AR headsets and a reduction of the field of view. We presented future technical improvements.
We hope to encourage other researchers to explore peripheral cues for HMDs using our open-source toolkit.

\bibliographystyle{abbrv-doi-hyperref}
\bibliography{zotero,manual}
\end{document}